\documentclass[manuscript]{aastex62}

\usepackage{placeins}
\usepackage{morefloats}
\usepackage{graphicx}

\newcommand{\kms}[0]{km~s$^{-1}$}

\shorttitle{MORPHOLOGY OF THE INNER LISM}
\shortauthors{Linsky, Redfield, and Tilipman}

\begin{document}
\title{THE INTERFACE BETWEEN THE OUTER HELIOSPHERE AND THE INNER LISM:
MORPHOLOGY OF THE LOCAL INTERSTELLAR CLOUD, ITS HYDROGEN HOLE,
STROMGREN SHELLS, AND $^{60}$Fe ACCRETION\footnote{Based on
observations made with the NASA/ESA
Hubble Space Telescope obtained from the Data Archive at the Space
Telescope Science Institute, which is operated by the Association of
Universities for Research in Astronomy, Inc., under NASA contract NAS
AR-09525.01A. These observations are associated with programs \#12475,
12596.}}

v%% Note that \email has replaced the old \authoremail command

\author{Jeffrey L. Linsky}
\affiliation{JILA, University of Colorado and NIST, Boulder, CO 
80309-0440, USA}

\author{Seth Redfield}
\affiliation{Astronomy Department and Van Vleck Observatory,
  Wesleyan University, Middletown, CT 06459-0123, USA}

\author{Dennis Tilipman}
\affiliation{Astrophysics and Planetary Sciences Department and JILA,
  University of Colorado, Boulder, CO 80309-0440}
 
\correspondingauthor{Jeffrey L. Linsky}
\email{jlinsky@jila.colorado.edu}



%\today

\begin{abstract}

We describe the interface between the outer heliosphere and the local
interstellar medium (LISM) surrounding the Sun. The components of the inner LISM
are the four partially ionized clouds [the Local Interstellar Cloud (LIC), 
G cloud, Blue cloud, and Aql cloud] that are in contact with the outer 
heliosphere, and ionized gas
produced by EUV radiation primarily from $\epsilon$~CMa. We
construct the three-dimensional shape of the LIC based on interstellar line 
absorption along 62 sightlines and show that in the direction of 
$\epsilon$~CMa, $\beta$~CMa, and Sirius~B 
the neutral hydrogen column density from
the center of the LIC is a minimum. We call this
region the ``hydrogen hole''. 
In this direction, the presence of Blue cloud absorption and the absence of LIC
absorption can be simply explained by the Blue cloud lying just
outside of the heliosphere. We propose that the outer edge of the 
Blue cloud is a
Str\"omgren shell driven toward the heliosphere by high pressures in
the H~II region. We 
find that the vectors of neutral and ionized helium flowing through
the heliosphere are inconsistent with the LIC flow vector, and that
the nearby intercloud gas is consistent with ionization by
$\epsilon$~CMa and other stellar sources
without requiring additional sources of ionization or 
million degree plasma. In the upwind direction, the heliosphere is
passing through an environment of several LISM clouds,
which may explain the recent influx of interstellar grains containing 
$^{60}$Fe from supernova ejecta measured in Antarctica snow.
 
\end{abstract}


%%\documentclass[12pt,preprint]{aastex}
%% manuscript produces a one-column, double-spaced document:
%\documentclass[manuscript]{aastex}
%% preprint2 produces a double-column, single-spaced document:
 %\documentclass[preprint2]{aastex}
%% Sometimes a paper's abstract is too long to fit on the
%% title page in preprint2 mode. When that is the case,
%% use the longabstract style option.
%% \documentclass[preprint2,longabstract]{aastex}

%%%%%%%%%%%111111111111111111111111111111111111111111111111111111111111

\section{Introduction}

\subsection{Properties of partially ionized warm clouds in the local
  interstellar medium}

The Sun is surrounded by patchy network of warm (5,000--10,000~K) 
partially ionized gas  
clouds extending out to a distance of about 15~pc in the local interstellar
medium (LISM). We have learned about the properties
of these gas clouds from high-resolution spectra of absorption lines 
produced by interstellar gas along lines of sight to nearby stars 
and from satellite measurements  
of interstellar gas flowing through the heliosphere. 
\cite{Crutcher1982} first reported that the radial velocities of
interstellar absorption lines in
the spectra of nearby stars are consistent with interstellar gas
flowing toward the Sun from the direction of the Scorpio-Centaurus 
Association. Later
investigations \citep[e.g.,][]{Lallement1992,Frisch2002} found
that the local interstellar gas flow has a number of velocity 
components with slightly different flow directions and speeds. 

\begin{figure}[t]
\begin{center}
\includegraphics[angle=90,scale=0.6]{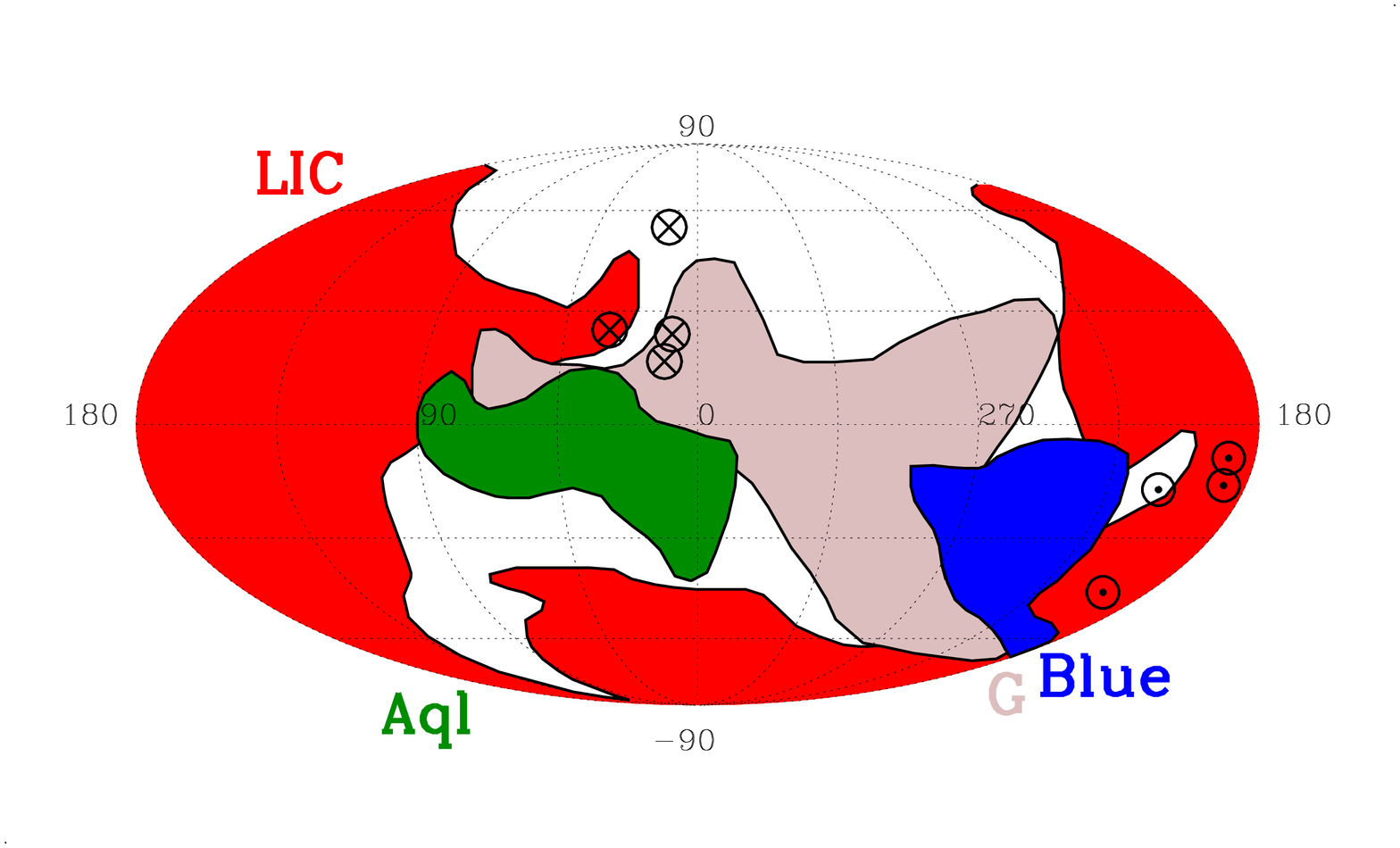}
\end{center}
\caption{Morphologies of the four partially ionized LISM clouds 
  that are in contact with the outer heliosphere. They are
  the LIC (red), which lies in front of $\epsilon$~Eri (3.2~pc), 
  the G cloud (brown), which lies in front of 
  $\alpha$~Cen (1.32~pc), the Blue cloud (dark blue), which lies in
  front of Sirius (2.64~pc), and the Aql cloud (green), which lies in front 
  of 61~Cyg (3.5~pc). The plot is in Galactic
  coordinates with the Galactic Center direction in the center. The
  upwind direction of the LIC velocity vector is
  indicated by the circled-cross symbol near $l=15^{\circ}$ and 
  $b=+20^{\circ}$, and the upwind directions of the other clouds have
  similar marks. The downwind
  directions are indicated by the circled-dot symbols.. A full map
  of all 15 LISM clouds is given by  
  \cite{Redfield2008}.\label{allclouds}}
\end{figure}

\cite{Redfield2008} subsequently identified 15 different velocity
components of warm interstellar gas located within 15~pc of the Sun
by analyzing interstellar absorption lines in {\em HST} spectra of 157 
nearby stars. The measured radial velocities along the lines of sight to
stars distributed within large solid angles allowed
\cite{Redfield2008} to determine velocity vectors for each of the 15 
clouds. Using these velocity vectors, \cite{Malamut2014} predicted accurate
radial velocities for the interstellar gas along the lines of sight to
all 34 stars observed in their new data set. 
This success in predicting accurate radial velocities for these new sight
lines demonstrated that these cloud vectors have accurate predictive
power. Figure~\ref{allclouds} shows the angular extent 
in Galactic coordinates of the four closest
interstellar gas clouds named LIC, G, Blue, and Aql.
The Local Interstellar Cloud (LIC) is so-named
because its angular extent covers nearly half of the sky, implying
that the Sun is located just inside of the LIC or possibly immediately outside.
The decision as to which option is more likely valid requires a
second data type --- 
measurements of interstellar gas flowing through the heliosphere.

Table~\ref{tab:inflow} compares the properties of the interstellar gas flowing
through the heliosphere with the parameters of gas located in the four nearest
interstellar clouds. The properties listed are the flow speeds 
($v_{\rm LISM}$) relative to the Sun, temperatures ($T$) inferred 
from the interstellar line widths, and the ecliptic longitude
($\lambda$) and latitude ($\beta$) of the flow. In the table, we 
list parameters for neutral and ionized helium
gas flowing through the heliosphere from the LISM as measured by four 
spacecraft: the {\em Extreme Ultraviolet Explorer (EUVE)},
{\em Interstellar Boundary Explorer (IBEX)}, {\em Ulysses}, and the 
{\em Solar TErrestrial Relations Observatory (STEREO)}. {\em EUVE}
measured the resonant scattering of solar EUV photons by
inflowing neutral helium atoms.
{\em IBEX} measured the direction of neutral helium
atoms that flow through the heliosphere without direction changing collisions. 
The resulting parameters obtained from four analyses of {\em IBEX} 
data listed in the table are in excellent agreement, indicating that the
inflow speed of LISM gas relative to the Sun's motion though the LISM
is about $v_{\rm LISM}$=26~km~s$^{-1}$ and the inflow direction is
given by ecliptic longitude $\lambda=75.5^{\circ}$ and ecliptic longitude of
$\beta=-5.2^{\circ}$. The most recent {\em IBEX} measurement of inflowing 
helium  \citep{Swaczyna2018} refers to the primary component, which is
separated from the ``warm breeze'' secondary component.

The inflow
direction of neutral helium observed by the {\em Ulysses} spacecraft and He$^+$
pick-up ions (PUI) measured with {\em STEREO} spacecraft data also agree with
$\lambda=75.5^{\circ}$. Helium pick-up ions are previously neutral
helium atoms that were ionized by EUV photons or charge-exchange processes
in the heliosphere, then
picked-up by the magnetic solar wind and gravitationally focused into
a cone in the downwind direction. \cite{Taut2018} have investigated
possible systematic errors in the analysis of the {\em STEREO}
He$^+$ pick-up ion data, but they found no significant change in $\lambda$
compared to the earlier results \citep{Mobius2015b}, except for a more
realistic assessment of the errors.
Since the analysis of {\em IBEX} data of neutral
helium results in a
tight coupling between $v_{\rm LISM}$, $\lambda$, and $\beta$
\citep{McComas2015,Mobius2015a,Mobius2015b}, independent measurements
of $\lambda$
from {\em Ulysses} and {\em STEREO} are essential in pinning down
accurate values for $v_{\rm ISM}$ and $\beta$. There have been a
number of studies concerning whether the inflow vector obtained from
{\em IBEX} observations of neutral helium is affected by
interstellar magnetic fields or by confusion with a second component 
of the inflowing
helium, called the ``Warm Breeze'' \citep{Kubiak2014},
but these possible effects appear to be very small \citep{Kubiak2016}

Included in Table~\ref{tab:inflow}  are the properties of the neutral 
hydrogen gas located in the nearby partially ionized LISM clouds.
\cite{Redfield2008} obtained these properties from their analysis of 
interstellar absorption lines in {\em HST} spectra of nearby stars. 
Also included in the table is a reanalysis 
of the flow vector through the LIC including 25\% more sightlines than 
were not available at the time of the 2008 paper. The addition of
these new sightlines produced only a slight change in the LIC flow
parameters. The LIC cloud provides the closest match to the inflow
parameters provided by {\em IBEX}, {\em Ulysses}, and {\em STEREO}, but
the match is not perfect. We will discuss this agreement or
disagreement in Section 6.

\cite{Slavin2008} computed a
model for the LIC with neutral hydrogen number density 
$n_{\rm  HI}=$ 0.19--0.20~cm$^{-3}$, electron density 
$n_e=0.007\pm0.01$~cm$^{-3}$ and temperature $T=6300$~K. \cite{Redfield2008}
found that the temperature of gas in the LIC is $7500\pm 1300$~K,
that the temperatures of the other clouds lie in the
range 5300--9900~K, and that their neutral hydrogen column densities lie in
the range $\log N_{\rm HI}=$ 17.2--18.8. The values of $n_{\rm HI}$ and $n_e$ 
in the other clouds are unknown, although the clouds are likely partially
ionized like the LIC. 

The absence of interstellar absorption at the predicted LIC velocity
in the direction of the Sun's motion implies that the Sun will leave
the LIC in less than 3000 years \citep{Redfield2008}. 
\cite{Frisch2013} and \cite{Frisch2015} 
proposed that the inflow direction has changed over the last 40 years,
suggesting that the heliosphere's environment is changing in our lifetime. 
\cite{Lallement2014}, however, argued against changes in the neutral helium
inflow direction based on a reanalysis of the {\em IBEX} data including
dead-time counting effects and the ecliptic longitude directions of
pick-up ions measured by {\em STEREO}. Their conclusion was supported by the
absence of any measurable change over a 20-year time span in the 
interstellar flow vector of neutral hydrogen measured by
the Solar Wind ANisotropies (SWAN) experiment on {\em SOHO} 
\citep{Koutroumpa2017}.

\subsection{What are the properties of the intercloud gas?}

The theoretical models of the interstellar medium proposed by \cite{Field1969}, 
\cite{McKee1977}, and \cite{Wolfire1995} decribe the interstellar gas as
consisting of three components: cold ($T\leq 50$~K) neutral and molecular gas,
warm neutral or partially ionized gas, and million-degree
low-density fully ionized plasma. These classical models assume that the three
components are each in thermal equilibrium and coexist in pressure 
equilibrium, but steady
state equilibrium is highly unlikely in the low density dynamic interstellar
medium where the time scales for ionization and recombination are 
on the order of $10^7$ years \citep{Chassefiere1986}.
The warm partially ionized gas clouds within a few parsecs of the Sun 
have properties roughly consistent with the warm component
predicted by the classical models, and dense cold molecular clouds are
observed typically by CO and H~I 21-cm emission. 
The nearest cold gas with a temperature of 15--30~K is the 
Leo cloud located 
at a distance between 11.3 and 24.3~pc from the Sun \citep{Peek2011}. 
However, numerical simulations by \cite{Berghofer2002}, which include supernova
explosions and realistic thermal and dynamic processes, predict a very 
wide range of densities and temperatures in the ISM but no pressure equilibrium
and no identifiable thermal phases.

The Sun is located in a low-density region called the Local Cavity
that extends more than 80~pc in all directions \citep{Frisch2011}. 
Inside of the Local Cavity are at least 15 partially ionized warm clouds
\citep{Redfield2008} and intercluster gas
that was originally assumed to be hot (roughly $10^6$~K), fully
ionized, and low density (roughly 0.005~cm$^{-3}$). 
This Local Hot Bubble model was supported by the predictions of
the classical models and observations of diffuse soft X-ray emission 
detected by rocket experiments and the {\em ROSAT} satellite. However,
the presence of hot gas in the Local Cavity is now challenged on the basis
of the following observational problems presented by \cite{Welsh2009}:

\begin{description}
\item[Solar wind charge exchange (SWCX) emission] The unexpected
  detection of X-ray emission from Comet Hyakutake \citep{Lisse1996} 
  led to the recognition that charge exchange reactions between solar 
  wind ions and neutral gas in the heliosphere can
  produce X-ray emission \citep{Cravens1997} that is
  similar to the emission produced by a million degree
  plasma. This result led to two different scenarios: (1) that roughly half
  of the observed diffuse X-ray emission in the Galactic plane is produced by 
  SWCX reactions inside the heliosphere with the other half produced
  by hot intercloud plasma
  \citep{Robertson2003,Galeazzi2014}, or (2) essentially all of the 0.75 keV
  emission in the Galactic plane is SWCX emission and there is no need
  for emission from a hot plasma except near the Galactic poles
  \citep{Snowden1994,Cox1998,Koutroumpa2009,Koutroumpa2012}.

\item[O~VI absorption] If hot gas were present in the Local Cavity,
  then interstellar absorption in the far-ultraviolet lines of O~VI 
  would indicate that intermediate temperature ($T\approx 300,000$~K) gas 
  is present where the hot gas comes in contact with cooler gas at the
  edges of the partially ionized warm gas clouds. O~VI
  absorption lines are detected in the circumstellar environment of hot
  stars and at high Galactic latitudes where there is hot gas in
  contact with the Galactic halo, but O~VI absorption is not detected in
  lines of sight towards stars within 58~pc of the
  Sun \citep{Barstow2010}. The intercloud gas must, therefore, be cooler than
  300,000~K yet still be mostly ionized so as to not show neutral
  hydrogen absorption.

\item[Pressure imbalance] If the diffuse X-ray emission were produced
  by hot plasma, then the inferred emission measure predicts a gas
  pressure $P/k$ = 10,000--15,000~cm$^{-3}$K \citep{Snowden2014b} 
  that is much larger than the gas pressure in the warm partially 
  ionized clouds like the LIC where 
  $P/k \approx 2500$~cm$^{-3}$K \citep{Redfield2008}. 
  While additional pressure terms
  (e.g., magnetic fields, cosmic rays, and ram pressure) may be
  important, the very large pressure difference argues against the presence
  of hot plasma at least in the Galactic plane.

\item[Upper limits on EUV line emission] Upper limits for diffuse
  high-temperature emission obtained by the {\em EURD}
  (Espectr\'ografo Ultra-violeta extremo para la Radiaci\'on Difusa) 
  satellite \citep{Edelstein2001} exclude significant emission
  from both $10^6$~K and intermediate temperature ($10^5$~K) gas in the 
  Local Cavity. Upper limits obtained with the {\em Cosmic Hot
  Interstellar Plasma Spectrometer (CHIPS)} satelite by \cite{Hurwitz2005}
  for diffuse emission of Fe lines, in
  particular the \ion{Fe}{9} 171.1~\AA\ line, are also inconsistent with the
  predicted emission from putative $10^6$~K thermal plasma in the Local Cavity.

\end{description}

Given these strong arguments against the presence of hot gas in the Local
Cavity except towards the Galactic poles, the gas located between the warm 
partially ionized clouds
(intercloud gas) and elsewhere within 80~pc of the Sun must be ionized
but not necessarily hot in order to be not detected as neutral gas. 
Upper limits on the non-SWCX  
X-ray emission requires that the intercluster gas must be
much cooler than $10^6$~K  or have a very low emission measure
as indicated by X-ray shadowing experiments \citep[e.g.,][]{Peek2011}
and by extreme ultraviolet spectroscopy \citep{Hurwitz2005}.

Various authors have proposed different
solutions to the intercloud gas problem by identifying different
sources of past and present ionization. \cite{Lyu1996} and
\cite{Breitschwerdt1999} proposed that
the intercloud gas is a recombining remnant of a 
past ionization event such as a supernova shock wave. In this
non-equilibrium plasma, the
degree of ionization can be far higher than the electron temperature of the
gas. This model is supported by the presence of young massive stars in
the nearby Scopius-Centaurus OB Association and the likely presence of
a previous group of massive stars that produced many supernova
explosions with the last supernova perhaps as recent as 0.5 Myr. 
\cite{Welsh2009} proposed a ``Hot-Top model'' in which there is no
hot gas except near the Galactic poles, but elsewhere the intercloud
gas is highly ionized with an electron temperature of about 20,000~K
in rough pressure equilibrium with the partially ionized warm clouds. 
An important source of ionization is the EUV radiation from
$\epsilon$~CMa, the brightest EUV source detected by the 
{\em Extreme Ultraviolet Explorer (EUVE)} satellite and other hot stellar
and white dwarf sources \citep{Vallerga1995, Vallerga1998}.
Among the nearby sources of EUV emission is Sirius~B, located only 
2.6~pc from the Sun.

\cite{Stromgren1939} showed that the EUV emission of hot stars
photoionizes the surrounding gas producing an HII region extending out to a
distance that defines a classical Str\"omgren sphere. Our model for
the intercloud gas near the Sun is a Str\"omgren sphere-like H~II
region photoionized primarily by $\epsilon$~CMa rather than a
recombining plasma, because the ionization state of the gas seen towards
$\epsilon$~CMa is modest (mostly singly-ionized atoms) and the EUV
radiation field is very strong. 

\subsection{Outline of this paper}

In this paper, we describe the properties of the partially
ionized warm gas clouds in the immediate neighborhood (within 4 pc) 
of the Sun and the intercloud gas present between these clouds. In a
subsequent paper, we will extend this analysis further into the LISM.
In Sections 2 and 3, we measure the size and shape of the LIC from 62 column
densities of \ion{D}{1}, \ion{Fe}{2}, and \ion{Mg}{2}. Section 4
describes the properties of Str\"omgren spheres of ionized gas
surrounding nearby hot stars and white dwarfs.  In Section 5, 
we identify the hydrogen hole in the LIC, which is
photoionized by EUV radiation primarily from $\epsilon$~CMa, and
propose that the Blue cloud in the direction of the hydrogen hole is
a Str\"omgren shell. In Section 6 we consider whether the flow vector
measured by {\em IBEX} and other satellites is inconsistent with the
LIC flow vector measured from interstellar absorption lines,
and, if so, what information can be gleaned from this inconsistency.
In Section 7, we
identify the gas clouds and intercloud components in the sightlines to
nearby stars, In Section 8 we propose that the recent measurement
in Antarctic snow of 
interstellar grains containing $^{60}$Fe from a supernova could be
explained by the inflow of dust grains from warm clouds in contact
with the heliosphere either at a continuous low level rate or during
an unusual event. Finally, we
list our conclusions and needed future work. 

%%%%%%%%%%22222222222222222222222222222222222222222222222222

\section{DATA ANALYSIS}

\subsection{Estimating distances to the edge of the LIC}

The procedure for estimating distances to the edge of the LIC is conceptually 
straightforward.  We assume that the LIC surrounds the heliosphere in
most directions and has 
constant neutral hydrogen density $n$(H~I). Therefore, the neutral
hydrogen column density $N$(H~I) divided by $n$(H~I) 
gives the path length of the absorbing medium, which in this case 
is simply the distance to the edge of the LIC along this line of sight
(LOS).  A fairly robust measurement of $n$(H~I)
exists for the immediate vicinity of the Sun, derived from (a) 
Lyman-$\alpha$ backscatter estimates \citep{Quemerais1994}, (b) 
{\it in situ} neutral helium measurements \citep{Gloeckler2004}, and (c)
\ion{He}{1}/\ion{H}{1} measurements from extreme-UV observations of local 
white dwarfs \citep{Dupuis1995}.  These diverse measurements indicate that 
$n($\ion{H}{1}$) \approx 0.2$~cm$^{-3}$ in the immediate 
vicinity of the Sun.  This density is slightly higher than the lower limit of 
$n($\ion{H}{1}) derived from dividing $N$(H~I) by the 
distance to the nearest observed stars, which has a maximum value of 
$\langle n($\ion{H}{1}$)\rangle \sim 0.1$~cm$^{-3}$ This difference
between $n($\ion{H}{1}) and $\langle n($\ion{H}{1}$)\rangle$
indicates that either the filling factor of warm gas inside the LIC and
presumably other clouds is less than unity or that 
the portion of the LIC very close to the Sun has a higher density 
than elsewhere in the LIC \citep{Redfield2008}.  In addition,
several sight lines to nearby stars (e.g., 61 Cyg, 
$\alpha$ CMi, and $\alpha$ Aql) limit the average value of 
$\langle n($\ion{H}{1}$)\rangle$ to $\leq 0.2$~cm$^{-3}$.

Hydrogen column densities, however, are difficult to measure towards nearby 
stars.  The strongest transition (e.g., Lyman-$\alpha$) is saturated, 
contaminated by airglow emission, and complicated by heliospheric and 
astrospheric absorption \citep{Wood2005}.  We, therefore, estimate $N$(H~I)
from other available atoms and ions in the following priority: 
\ion{D}{1}, \ion{Fe}{2}, and \ion{Mg}{2}.  Deuterium is an excellent tracer 
of hydrogen due to its tight abundance ratio within the Local Bubble, 
D/H$ = 15.6 \pm 0.4 \times 10^{-6}$ \citep{Linsky2006}.  Both \ion{Fe}{2} and 
\ion{Mg}{2} are the dominant ionization stages for these elements, and despite 
significant depletion onto dust grains, they have relatively tight and narrow 
correlations with hydrogen.  Also, these three ions are commonly observed.  
Of the 79 sight lines assigned to the LIC on the basis of their radial
velocities being consistent with the LIC velocity vector 
\citep{Redfield2008}, 64 (81\%) have observations in one or more of these 
ions.  The remaining LIC sight lines were detected in \ion{Ca}{2} alone.  Most 
measurements were taken from compilations of observations of these ions from 
\citet{Redfield2004a} for \ion{D}{1} and \citet{Redfield2002} and 
\citet{Malamut2014} for \ion{Fe}{2} and \ion{Mg}{2}. 

We use LIC sight lines with measurements of \ion{D}{1}, 
\ion{Fe}{2}, and \ion{Mg}{2} to empirically determine the appropriate 
conversion to hydrogen column densities.  
\citet{Redfield2008} calculated the depletion of 
\ion{Fe}{2} and \ion{Mg}{2} inside the LIC based on 12 sight lines with both 
\ion{D}{1} and \ion{Fe}{2}, and 21 sight lines that have both \ion{D}{1} and 
\ion{Mg}{2} observations.  The weighted mean value of \ion{Fe}{2}/\ion{H}{1}, 
where the hydrogen column density is calculated from \ion{D}{1} and the 
well-determined D/H ratio given above, is 
$\langle$\ion{Fe}{2}/\ion{H}{1}$\rangle = 2.14^{+0.61}_{-0.48} \times 10^{-6}$, 
and for magnesium, 
$\langle$\ion{Mg}{2}/\ion{H}{1}$\rangle = 3.6^{+2.8}_{-1.6} \times 10^{-6}$.  
The errors include the error in the D/H ratio, as well as the standard 
deviation in the distribution of measurements, which would include any 
depletion or ionization variations within the LIC.  

\begin{figure}
\plotone{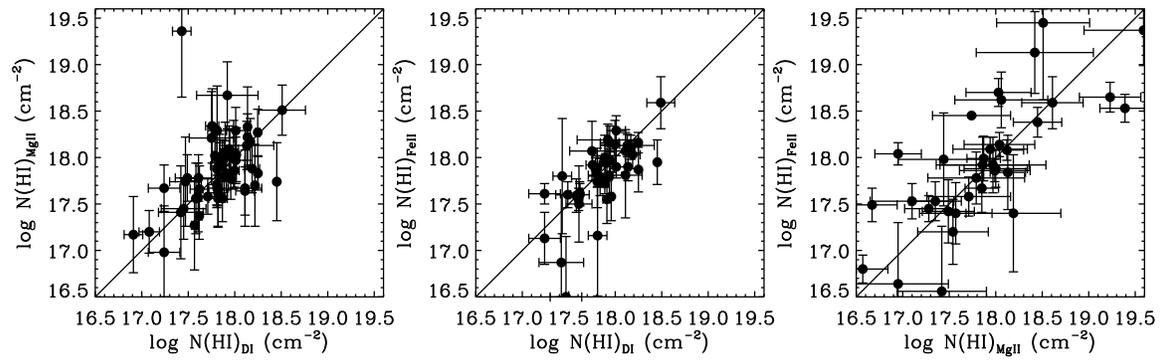}
\caption{Comparison of \ion{H}{1} column densities based on observed 
\ion{D}{1}, \ion{Mg}{2}, and \ion{Fe}{2} column densities.  
\label{fig:columnest}}
\end{figure}

Figure~\ref{fig:columnest} shows a comparison of the estimated \ion{H}{1} 
column densities based on observations of the three ions (\ion{D}{1}, 
\ion{Mg}{2}, and \ion{Fe}{2}).  Typical errors in the log \ion{H}{1} 
column density 
estimate based on \ion{D}{1} are only $\sim$0.09, while for \ion{Mg}{2} and 
\ion{Fe}{2} they are 0.35 and 0.27, respectively.   Note the tight 
correlation and small errors associated with \ion{D}{1}.  For all three 
comparison plots, $\sim$77\% of the data pairs predict consistent
values for $N$(H~I) pairs within 1$\sigma$.  While there is a large 
dispersion in the comparison of $N$(H~I) for \ion{Mg}{2} and \ion{Fe}{2}, 
they still seem to be relatively 
good proxies for the \ion{H}{1} column density, provided there is a sensible 
estimate of the elemental depletion.  

Other ions were investigated, but they suffer more seriously from 
ionization and depletion effects.  For example, \ion{Ca}{2} is not the 
dominant ionization stage of calcium in the LISM, but makes up only 1.6\% of 
the gas phase calcium, whereas 98.4\% is expected to be in \ion{Ca}{3} 
\citep{Slavin2008}.  For this reason, the \ion{Ca}{2}/\ion{H}{1} ratio varies 
much more significantly than the corresponding ratios for \ion{Mg}{2} and 
\ion{Fe}{2}, and does not provide a useful means of estimating 
the hydrogen column density.

The magnitudes of the errors in the abundance ratio to hydrogen (X/H) clearly 
show that deuterium measurements are most desirable, typically followed by 
\ion{Fe}{2}, and then by \ion{Mg}{2}.  Both the \ion{Fe}{2} and \ion{Mg}{2} 
measurements are given for all LIC sight lines in Table~\ref{tab:licmem}.  
Along six lines of sight, both Fe~II and Mg~II were measured and are
listed in the table. In all 
six cases, the two ions lead to $N$(H~I) estimates that 
agree to within 3$\sigma$.  The estimates of $N$(H~I) used in the
subsequent analysis are typically the ones with higher accuracy, which 
for these sight lines without D~I data were obtained from  
\ion{Fe}{2} in all cases.  Of the 67 sight lines with observations of the LIC 
in one of these three ions, 34 (51\%) have \ion{D}{1} observations, 19 (28\%) 
have \ion{Fe}{2} but no \ion{D}{1}, and 14 (21\%) were observed in \ion{Mg}{2} 
alone.  Table~\ref{tab:licmem} identifies each sight line, and which ion was 
used in the \ion{H}{1} calculation.  We assume 
$n($\ion{H}{1}$) = 0.2$~cm$^{-3}$ to calculate the distance from the Sun to 
the edge of the LIC. 

Sight lines in which the relative error in the distance to the LIC edge 
[$\sigma(d_{\rm edge})/d_{\rm edge}$] is of order unity (i.e., $>$0.9) were 
removed from the sample, as they do not provide any significant constraint on 
the distance to the LIC edge and result typically from large errors 
in the measured ion column density because of saturation or the difficulty in 
establishing the velocity component structure of blended profiles.  Only 3 
sight lines were removed ($\sim$4\% of the sample) for this reason.  
In addition, sight 
lines for which the distance to the LIC edge is estimated at more then 5 
standard deviations from the median value were also removed.  Again, it is 
possible that these are due to erroneously large column density measurements 
(in all cases they were larger than the median value) due to saturation or 
blending.  Five measurements were thus removed ($\sim$7\% of the sample), 
with three of them observed in \ion{Fe}{2} and two in \ion{Mg}{2} (which had 
complimentary \ion{Fe}{2} observations that could be used).  No targets 
observed in \ion{D}{1} were removed for these reasons.  
We note that three of these targets are roughly in the same part of the sky
(l=47$^{\circ}$--57$^{\circ}$, b=27$^{\circ}$--32$^{\circ}$): 
72 Her, 99 Her, and HD157466.  They all have nearly saturated absorption 
lines, leading to the large column densities and high path length measurements 
with large errors.  This part of the sky is near the upwind direction and is 
complicated by the presence of several LISM clouds 
\citep[see Fig. 19 in][]{Redfield2008}.  
Also, other LIC absorbers in the vicinity (e.g., $\gamma$ Ser, $\alpha$ Oph, 
$\delta$ Cyg, $\iota$ Oph) show relatively high column densities.  Three of 
these sight lines were observed only in \ion{Ca}{2} ($\alpha$ Oph, 
$\delta$ Cyg, $\iota$ Oph), but have among the highest values of $N$(H~I) 
detected in the LISM \citep[see Fig. 7 in][]{Redfield2002}. 
While $\gamma$ Ser and $\alpha$ Lyr are both within 5$\sigma$ of the 
mean LIC path length, both have observed values 
that are 2.7$\sigma$ from the predicted value based on our morphological 
model.  Clearly, something complex and interesting is occurring in this 
region of the sky.

A significant number of the \ion{D}{1} measurements were made without full 
knowledge of the velocity component structure.  If more than one absorbing 
cloud is present along the line of sight, this can lead to overestimated 
column densities if the profile is modeled as a single component.  Due to the 
strong thermal broadening of \ion{D}{1} \citep{Redfield2004b}, multiple velocity
components are not immediately obvious and thus require observations at 
high resolution of heavier ions such as \ion{Fe}{2} or \ion{Mg}{2} to
infer a more accurate value of $N$(H~I).  Because 
the contributions of such possible systematic errors are not included in the 
\ion{D}{1} column density estimate, we treat the estimates derived from these 
sight lines with caution.  Of the 34 \ion{D}{1} measurements, this effects 11 
(32\%) of them ($\tau$ Cet, $\delta$ Eri, EP Eri, HR1925, HR8, DX Leo, PW And, 
SAO136111, V471 Tau, HR1608, and Feige 24). Future short observations 
of \ion{Fe}{2} or \ion{Mg}{2} could easily resolve the component 
structure and greatly improve 
the accuracy of the \ion{D}{1} analysis \citep[e.g.,][]{Malamut2014}.

%%%%%%%%%%%%%%%%3333333333333333333333333333333333333333333333

\section{Morphology of the Local Interstellar Cloud (LIC)}

Motivated by a significant increase in the number of observations of
the LISM, we present a revised analysis of the three-dimensional morphology 
of the interstellar material that directly surrounds our solar system, 
the Local Interstellar Cloud (LIC). We follow the same procedure outlined 
in \citet{Redfield2000}, which involves fitting a series of spherical 
harmonics to the estimated distances to the edge of the LIC given 
in Table~\ref{tab:licmem}. As in \citet{Redfield2000}, we fit the data 
to 9 basis functions (e.g., $l = 0, 1, 2$). We assume a homogeneous and 
constant density cloud in order to estimate the distance to the edge of the 
LIC from our column density measurements. In Section 6 we will
test the validity of these assumptions. With a large enough sample and 
corresponding high orders of spherical harmonics, any arbitrary closed 
surface can be characterized with this technique. 
The best fitting amplitude for each
spherical harmonic basis function is determined using a least-squared 
minimization routine. 

There are two significant changes in our analysis compared to 
\citet{Redfield2000}. First is the quantity and quality of the input 
data. In order to have a sufficient number of data points, 
\citet{Redfield2000} used hydrogen column density measurements derived 
from observations using {\it HST}, {\it EUVE}, and ground-based 
\ion{Ca}{2}. This resulted in a sample comprised of 32 measurements. In 
the current analysis, we limit ourselves to a homogenous sample of only 
the highest quality column density measurements, derived from 
{\it HST} spectra. Given the significant increase in LISM measurements 
\citep[e.g.,][]{Redfield2002, Redfield2004a, Malamut2014}, our current 
sample includes 62 measurements.

The second significant difference with the \citet{Redfield2000} analysis, 
is our improved knowledge of the hydrogen volume density in 
the LISM. \citet{Redfield2000} used a value of $n_{\rm HI} =0.1$~cm$^{-3}$, 
based on limits toward the nearest stars. However, we now estimate the 
neutral hydrogen density to be closer to $n_{\rm HI} = 0.2$~cm$^{-3}$ 
\citep{Slavin2008}. Therefore, our estimates for the size of the LIC will be 
roughly a factor of two smaller than in \citet{Redfield2000} as
distances are inversely proportional to the assumed
neutral hydrogen number density.

\begin{figure}
\includegraphics[angle=0,width=5.9in]{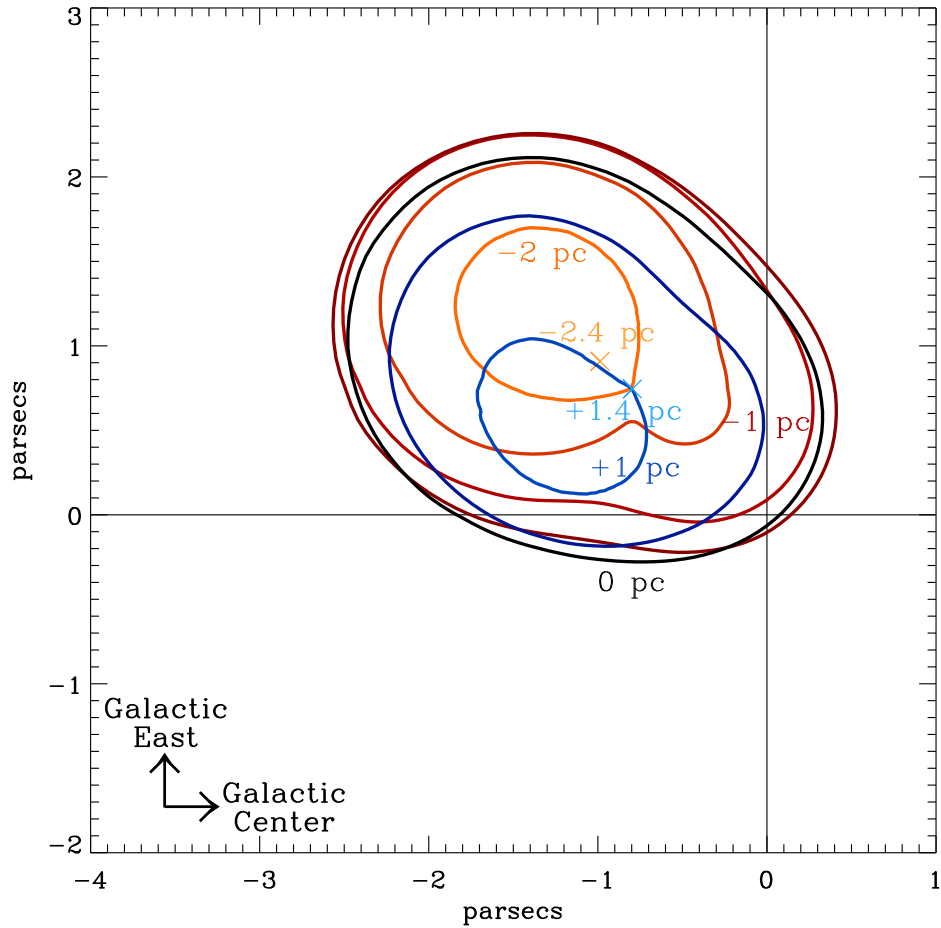}
\caption{Contour map of LIC as viewed from North Galactic Pole. The
  Sun is located at the origin (0,0). Red contours are cuts above the
  LIC center parallel to the Galactic plane, and the blue contours
  are cuts below. The X symbols indicate the locations where the
    edge of the LIC is furthest above and below the plane of the figure.
\label{fig:licconngp}}
\end{figure}

\begin{figure}
\includegraphics[angle=0,width=5.9in]{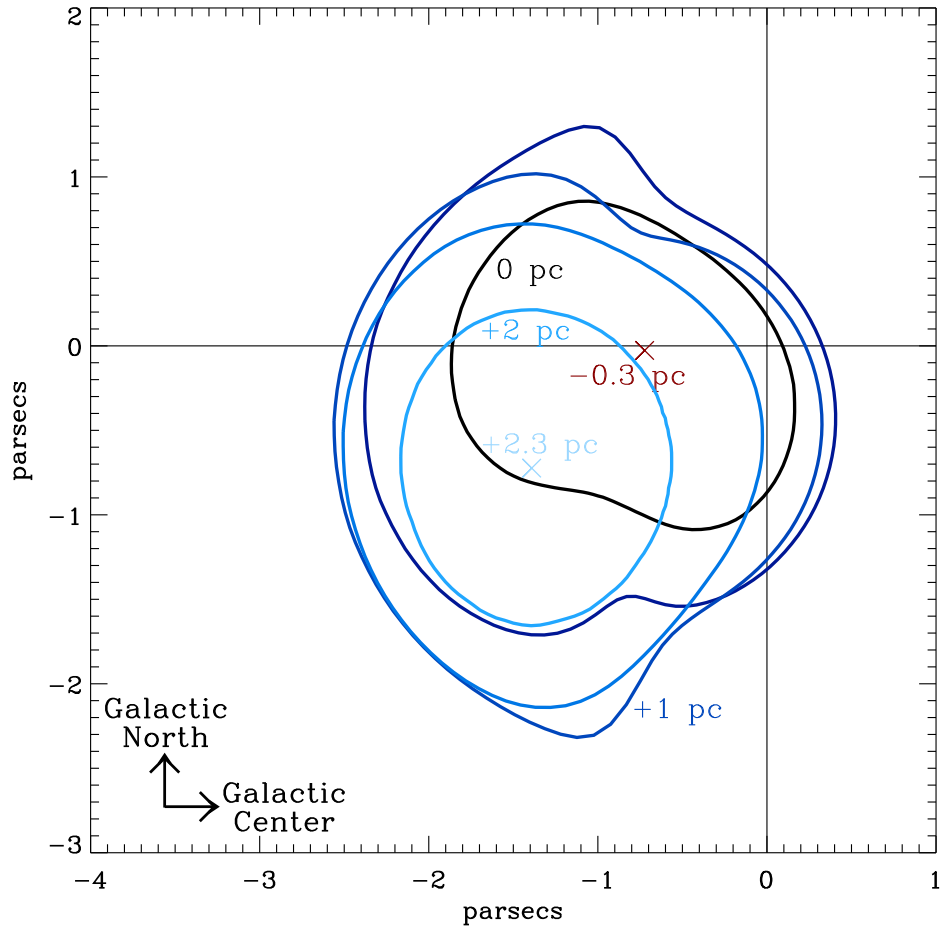}
\caption{Contour map of LIC as viewed from Galactic East. \label{fig:licconge}}
\end{figure}

\begin{figure}
\includegraphics[angle=0,width=5.9in]{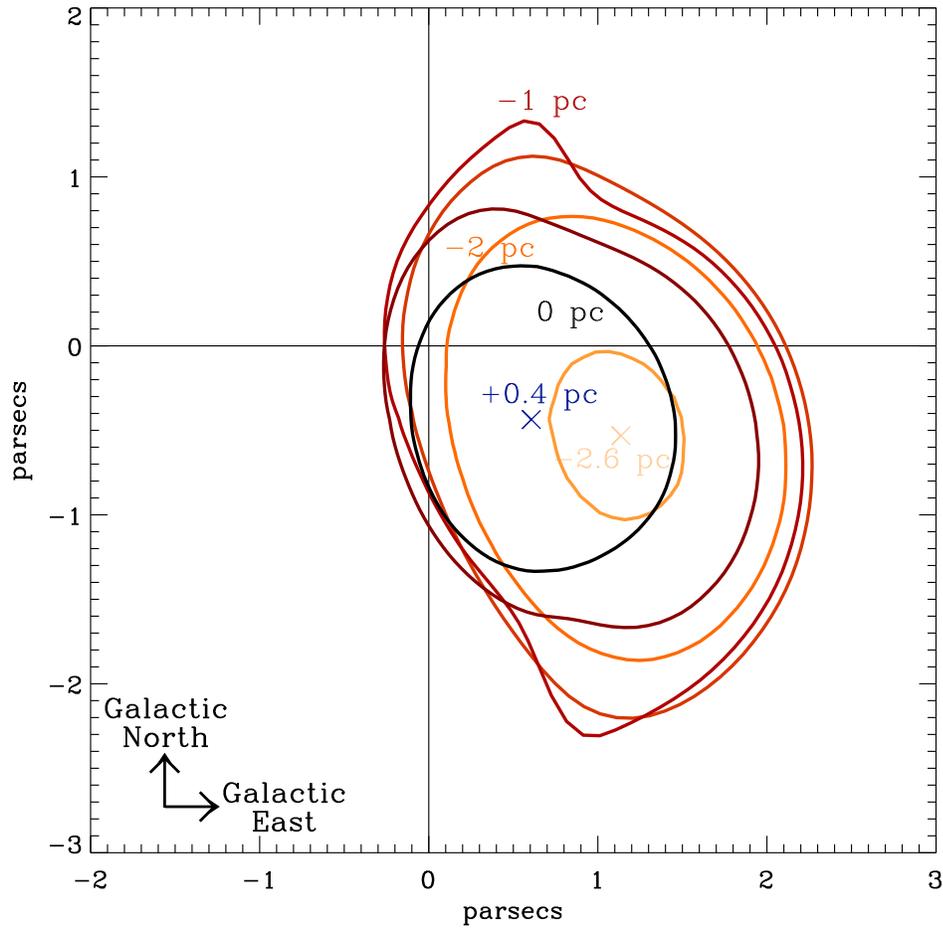}
\caption{Contour map of LIC as viewed from the Galactic Center. 
\label{fig:liccongc}}
\end{figure}

The resulting parameters of the best fit spherical harmonics model is 
given in Table~\ref{tab:licfit}. Like the original model in  
\citet{Redfield2000}, the dominant harmonic is the spherical $l=0$ 
component indicating that to first-order the LIC can be characterized 
as a quasi-spherical closed volume. However, the contributions of the 
additional spherical harmonic orders lead to significant departures 
from a pure sphere. Contours of the best fit three-dimensional model 
are shown in Figures~\ref{fig:licconngp}--\ref{fig:liccongc}. The 
shape is much better constrained than in the \citet{Redfield2000} model, 
although the general geometry is not all that different. In particular, 
the observations continue to support the interesting conclusion that 
the Sun is located very near to the edge of the LIC, indicating that 
in as little as $\approx$3000 years, the Sun will pass out of the LIC 
and into a different interstellar environment.

\begin{figure}[t]
\begin{center}
\includegraphics[scale=0.6]{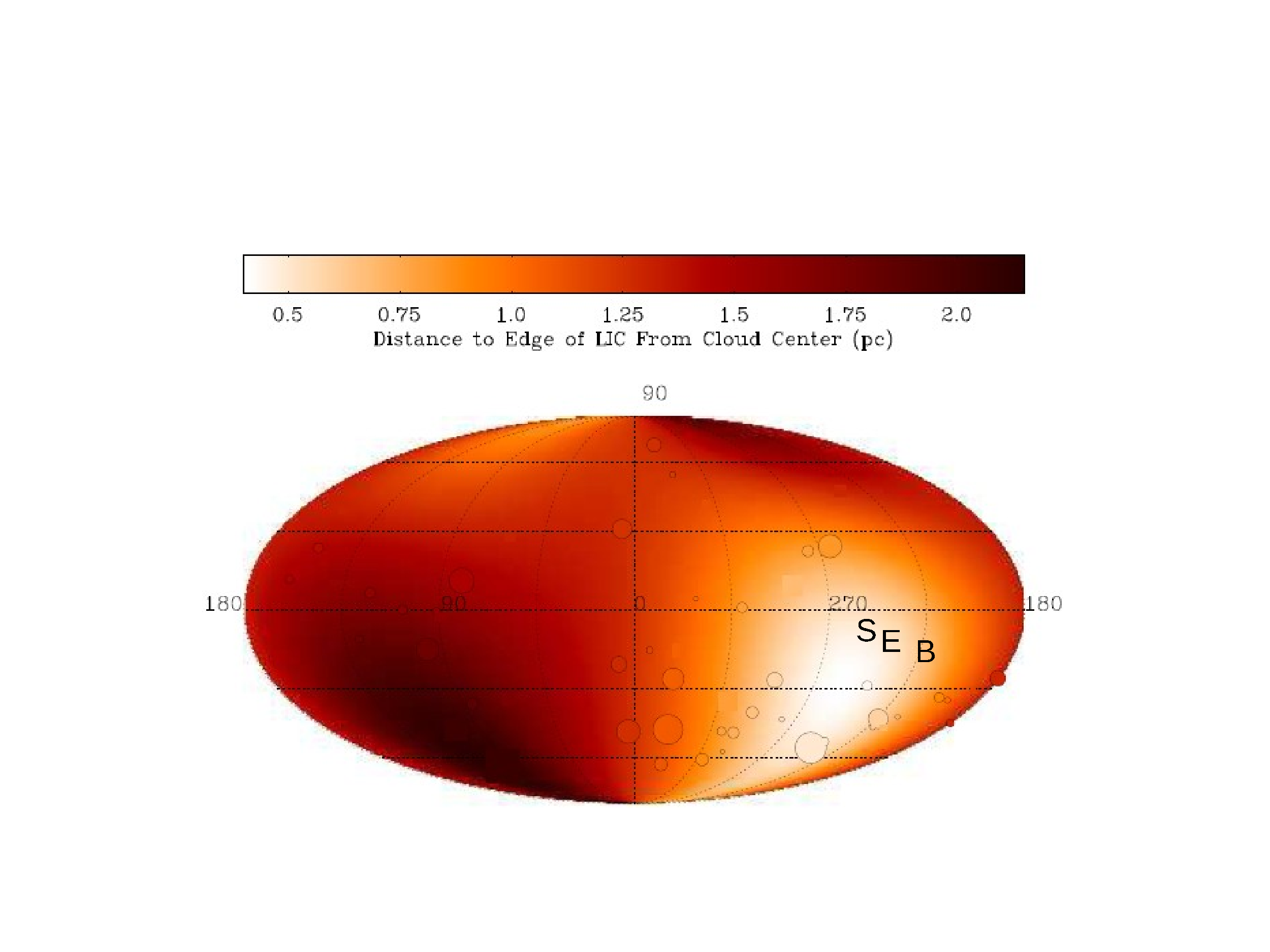}
\end{center}
\caption{Distances to the edge of the LIC from its geometric center
  computed from $N$(H~I) along lines if sight to 62 nearby stars. 
  The symbols are S for the Galactic coordinates of
  the hot white dwarf Sirius~B, E for $\epsilon$~CMa, and B for
  $\beta$~CMa. Faint circles indicate the locations of stars
    near the hydrogen hole. \label{LICfromCenter}}
\end{figure}

The geometry of the LIC can also be visualized in Figure~\ref{LICfromCenter}, 
where the shading indicates the distance to the edge of the LIC from 
the geometric center of the LIC (as given 
in Table~\ref{tab:licfit}). 
%Overplotted are the symbols that indicate 
%the sight lines used to generate the model and a shading interior to 
%the symbol indicate the observed distance so that it can be compared 
%to the predicted value around the symbol. 
The reduced $\chi^2$ is 
relatively high, indicating that there are significant departures 
from this relatively simple model. In particular, out of 62 
total data points, 26 (42\%) are within 1$\sigma$, 38 (61\%) within 2$\sigma$, 
and 54 (87\%) within 3$\sigma$. The most discrepant sight lines 
are toward $\eta$ Ari (for which the model predicts a larger distance than 
the observations imply) and $\tau^6$ Eri (for which the model predicts 
a much closer 
distance than the observations imply). Some of the discrepancies could 
be explained by misidentifications or blends of multiple components as 
singular LIC absorption features. However, it is likely that our simple 
assumptions of homogeneity and constant density are, unsurprisingly, 
not completely realistic. While, by and large, the structure is well 
characterized in this way, there are likely to be regions where there 
are significant departures in homogeneity or density that lead to 
discrepancies between the model and observations.

%%%%%%%%%%%444444444444444444444444444444444444444444444444444444444

\section{What is the Ionization Source for the Intercloud Gas near the Sun?}

The brightest nonsolar source of extreme-UV radiation detected by the 
EUVE satellite was the B2~II star $\epsilon$~CMa ($d$=124~pc) with
an intrinsic ionizating flux of about $2.7\times10^{46}$~s$^{-1}$  
\citep{Vallerga1995}. This flux estimate includes a correction for 
absorption by a hydrogen column density,  $N$(H~I)=$9\times 10^{17}$ cm$^{-2}$,
along the line of sight to the star. 
If the \cite{Gry1995} estimate of $N$(H~I)$<5\times 10^{17}$ cm$^{-2}$
is more realistic, then the intrinsic ionizing flux from $\epsilon$~CMa 
will be larger. The next brightest EUV source is 
$\beta$~CMa (B1 II-III; $d$=151~pc), followed by many hot white dwarfs located
inside of the Local Bubble \citep{Vallerga1998}. 
The total ionization rate of 33 hot white
dwarfs measured by EUVE is $\sim 1.6\times 10^{45}$ photons ~s$^{-1}$
\citep{Welsh2013}, which is more than a factor of 10 times smaller than the 
ionizing flux from $\alpha$~CMa. 

In a classic paper, \cite{Stromgren1939} showed that the EUV radiation
($\lambda < 912$~K) from a hot star completely ionizes hydrogen in its 
surrounding volume (called a Str\"omgren sphere) out to a distance now called 
the Str\"omgren radius where 
the build up of neutral hydrogen opacity absorbs the photoionizing radiation,
producing a narrow partially ionized shell surrounded by neutral hydrogen gas.
In this paper, Str\"omgren developed a simple model assuming that
the hot star is located in a constant density environment in which 
flows are ignored and 
photoionization of hydrogen is balanced by recombination in a steady state.
In this case, the radius of the classical Str\"omgren sphere is 

\begin{equation}
R^3=3 (dN_i/dt) /(4\pi\alpha n_in_e),
\end{equation}
where $dN_i/dt$ is the number of ionizing photons per second and $n_i$ and
$n_e$ are the number densities of ions and electrons inside of the
Str\"omgren sphere, and $\alpha$ is the recombination factor 
\citep{Harwit1988}. For relatively soft stellar radiation such as from
$\epsilon$~CMa where most of the ionizing radiation is in the
504--912~\AA\ band, hydrogen inside of the Str\"omgren sphere 
will be fully ionized and helium mostly
neutral. When the radiation field
is harder with significant radiation as wavelengths shortward of
the 504~\AA\ photoionization edge of He$^0$ or the 228~\AA\ photo-ionization
edge of He$^+$, as is the case for very hot white dwarfs such as G191-B2B
and HZ~43, then helium will be either singly or doubly ionized. 
\cite{Tat1999} estimated the sizes of Str\"omgren spheres around
hot white dwarfs in the Local Cavity using the classical
Str\"omgren sphere model. This model has been extended to include 
dust opacity, clumpiness, diffuse radiative transfer, and dynamics
\citep[e.g.,][]{Yorke1986}. \cite{McCullough2000} computed modified
Str\"omgren sphere models for the case of a hot star embedded in 
a larger ionized cavity. Depending on the location of the hot
star in the cavity, the H~II region around the star is no longer a sphere.
Rather, the H~II
region produced by the hot star is larger than for the classic case
because the surrounding gas is not neutral and the two H~II regions 
can merge. Since Sirius~B resides inside of
the H~II region produced by $\epsilon$~CMa, we refer to the H~II
region near Sirius~B as an ``extended H~II region''.

The electron density in the line of sight to $\epsilon$~CMa is unknown,
but if it is about 0.01 cm$^{-3}$, then the radius of the star's
Str\"omgren sphere equals the distance to the star (130 pc). This is
consistent with the conclusion by \cite{Welsh2013} that
$\epsilon$~CMa is the primary source responsible for the ionization of the
local ISM. They found that the volumetric
filling factor of the classical Str\"omgren spheres of all 33 of the hottest
white dwarfs (excluding Sirius~B) in the Local Cavity is less than 6\% 
and that none of these hot white dwarfs are close enough to the Sun 
to influence the local ionization. 

We next consider whether Sirius~B could be an important
local ionization source given its short 2.6~pc distance from the heliosphere. 
Fitting the {\em HST} spectrum of Sirius~B with a non-LTE model
atmosphere, \cite{Barstow2005} obtained 
the stellar parameters $T_{\rm eff}=25,193$~K, $\log g=8.556$, and radius
0.0081 solar. Martin Barstow (private communication) kindly computed the flux
shortward of 912~\AA\ for this model as $9.4\times 10^{39}$~photons~s$^{-1}$.
The radius of a classical Str\"omgren sphere for this photon flux is
0.25~pc for an assumed $n_e=0.1$ cm$^{-3}$ \citep{Redfield2008b} 
or 1.14~pc for an assumed
$n_e=0.01$~cm$^{-3}$. These calculations are for an isolated Str\"omgren
sphere surrounded by neutral hydrogen, but Sirius~B is embedded in 
the large H~II region ionized
by $\epsilon$~CMa, and the physical conditions of interstellar
gas near Sirius and the Sun are controlled by stellar radiation
in the 504--912~\AA\ region and by hot white dwarfs
at shorter wavelengths. 

Table~\ref{tab:composition} lists the stars within 16~pc of the Sun for 
which neutral hydrogen column densities through the nearby clouds have 
been measured
by \cite{Redfield2008} and by \cite{Malamut2014}. We list the distances 
through neutral hydrogen gas, $\Delta d$(neutral), along the 
lines of sight (LOS) through the identified clouds, assuming that the 
neutral hydrogen density is the same as that measured for the LIC 
($n_{HI}=0.2$ cm$^{-3}$) by \cite{Slavin2008}. We presume that 
the remaining path length $\Delta d$(ionized) is filled by 
ionized gas, but we revist this assumption when we discuss the G cloud
sight line in Section~7.

We note that the thickness of the partially ionized outer shell 
of a Str\"omgren sphere is $\delta=(n_{HI}\sigma)^{-1}$, where 
$\sigma\approx 10^{-17}$~cm$^2$ \citep{Harwit1988} is the 
hydrogen-ionization cross section for EUV photons near 912~\AA.
Thus for $n$(H~I)$\approx 0.2$~cm$^{-3}$, the
Stromgren shell thickness is $\delta\approx 0.2$~pc. Filamentary warm 
clouds like the Mic Cloud could have roughly this thickness. 

The incident EUV radiation from $\epsilon$~CMa should produce higher
ionization inside H~II regions and Str\"omgren shells. 
\cite{Gry1995} found interstellar absorption by Si~III and C~IV 
  at the predicted LIC velocity in the sight line to $\epsilon$~CMa,
  although the C~IV absorption could be in the stellar wind. 
  \cite{Dupin1998} also found C~IV absorption at the LIC cloud
  velocity in the sight line to $\beta$~CMa. Since both sight lines 
  pass through Str\"omgren shells, there is strong evidence for higher
  ionization in these shells.

In Table~\ref{shells}, we compare the thicknesses of the LIC and Blue clouds
seen in the directions of $\epsilon$~CMa and Sirius. The cloud thicknesses
in these directions are consistent with their being Str\"omgren
shells. It is likely that the outer edges of all clouds facing the EUV
radiation from $\epsilon$~CMa are Str\"omgren shells.

%%%%%%%%%555555555555555555555555555555555555555555555555555555

\section{The Hydrogen Hole and Blue Cloud have a common origin}

Figure~\ref{LICfromCenter} shows the distance to the edge of the LIC
from its geometric center computed from the measured values of 
$N$(H~I) along the lines of sight to 
nearby stars. The region within Galactic longitude 
$225^{\circ} \leq l \leq 290^{\circ}$ and Galactic latitude 
$-60^{\circ} \leq b \leq +10^{\circ}$ shows
very low H~I column densities corresponding to a skin depth of
$<0.5$~pc from the geometric center of the LIC.
We call this region the ``hydrogen hole''. The
Galactic coordinates of $\epsilon$~CMa, $\beta$~CMa, and Sirius as seen
from the center of the LIC all lie within this region. The location of the
hydrogen hole is consistent with the coordinates of these three strong
sources of ionizing radiation that apparently shape the morphology of
the LIC in the $\epsilon$~CMa direction.
\cite{Welsh1999} refer to the low hydrogen
column density along the lines of sight to $\epsilon$~CMa and
$\beta$~CMa as an interstellar tunnel or local chimney that extends 
beyond these stars to the Galactic halo. 

\begin{figure}
%Figure 7
\includegraphics[angle=0,width=5.9in]{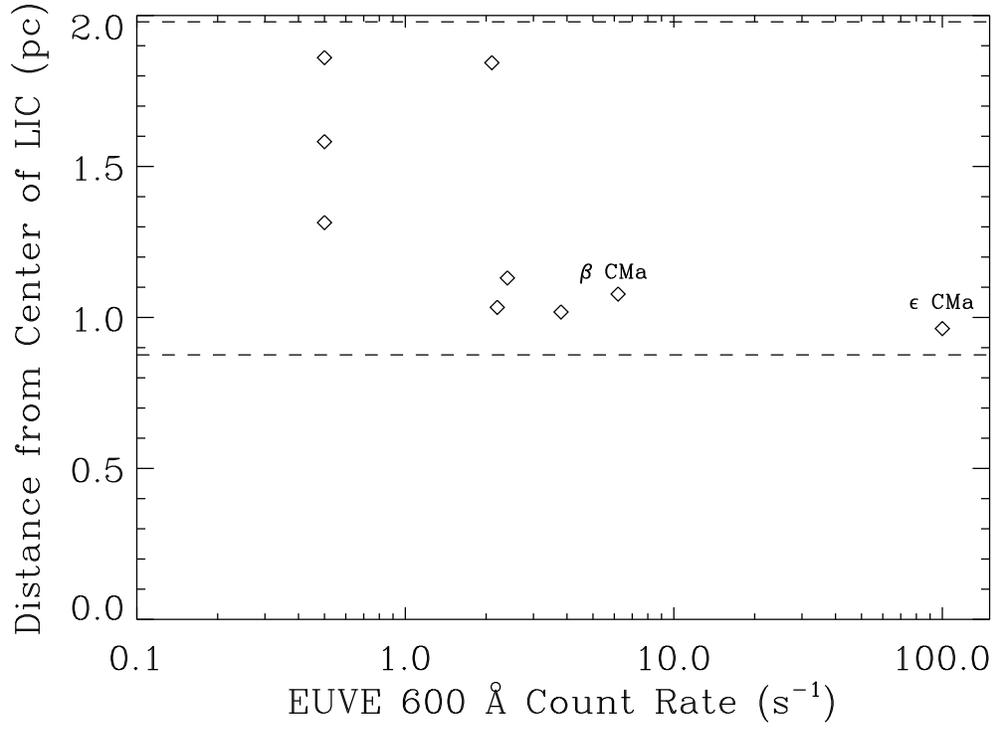}
\caption{Distance from geometric center of the LIC to its edge 
along the lines of sight to the brightest EUV targets observed by {\em EUVE}. 
\citep{Vallerga1995}. The x-axis data are count rates measured
through the wide band Tin filter centered near 600~\AA.
The sight lines to the brightest EUV sources
($\epsilon$~CMa at the far right and $\beta$~CMa) have among the
shortest path lengths to the center of the LIC. The dashed lines
indicate the minimum and maximum path lengths from the geometric
center of the LIC to its edge.
\label{fig:dliceuve}}
\end{figure}

Figure~\ref{fig:dliceuve} shows the distance in pc from the geometric center of 
the LIC to its edge along the sightlines to the 10 brightest EUV sources 
observed by the {\em EUVE} spacecraft \citep{Vallerga1995}. The lines
of sight to the brightest EUV 
source $\epsilon$~CMa and four of the five next brightest sources all have
the shortest distances from the center of the LIC edge to its edge,
which is consistent with EUV
radiation sculpting the LIC by photoionizing neutral hydrogen.

We now consider what type of gas lies immediately outside of the heliosphere in
the direction of the hydrogen hole. Three possibilities are
(i) a very thin layer of partially ionized LIC cloud gas, (ii) partially 
ionized gas of another cloud, or (iii) fully ionized 
hydrogen gas (H~II region). A test for the 
first two possibilities would be the detection of Lyman-$\alpha$ 
absorption in the hydrogen wall near the Sun where the
inflowing neutral hydrogen atoms from a partially ionized cloud (LIC
or another cloud) charge-exchange with the outflowing protons in the 
solar wind leading
to a pile-up of heated, red shifted, neutral hydrogen atoms. The result
would be
Lyman-$\alpha$ heliospheric absorption that is red shifted relative to the
inflowing neutral hydrogen. Table~\ref{tab:outside} summarizes the data for
those stars
located inside of or near the region of minimum $N$(H~I) that have
detected or nondetected solar hydrogen wall absorption. The data fall
into three groups detailed in Sections 5.1 to 5.3.

\subsection{Stars well inside of the hydrogen hole} 
The nine stars with lines of sight that traverse the hydrogen hole  
all show an absorption component at the predicted Blue
cloud radial velocity and, except for $\epsilon$~CMa, 
no absorption component at the predicted
LIC radial velocity. Three nearby stars, $\zeta$~Dor,
HR~2225, and HR~2882, also do not show solar hydrogen wall
absorption that would indicate the absence of neutral hydrogen in the
interstellar gas immediately outside of the heliosphere. For the other six
stars there are no high-resolution Lyman-$\alpha$ spectra needed
to determine whether or not they show solar hydrogen absorption.
However, all nine stars
are located in the downwind direction of the LIC flow where it
difficult to detect solar hydrogen wall absorption \citep{Wood2005}.
Therefore, immediately outside of the heliosphere in the hydrogen hole
direction there could be either Blue cloud or H~II gas. We propose that
Blue cloud gas is in contact
with the outer heliosphere along these lines of sight,
and that ionized hydrogen gas is located outside of the Blue cloud
where it receives unattenuated EUV radiation from $\epsilon$~CMa and Sirius.

\begin{figure}
\includegraphics[angle=0,width=7.0in]{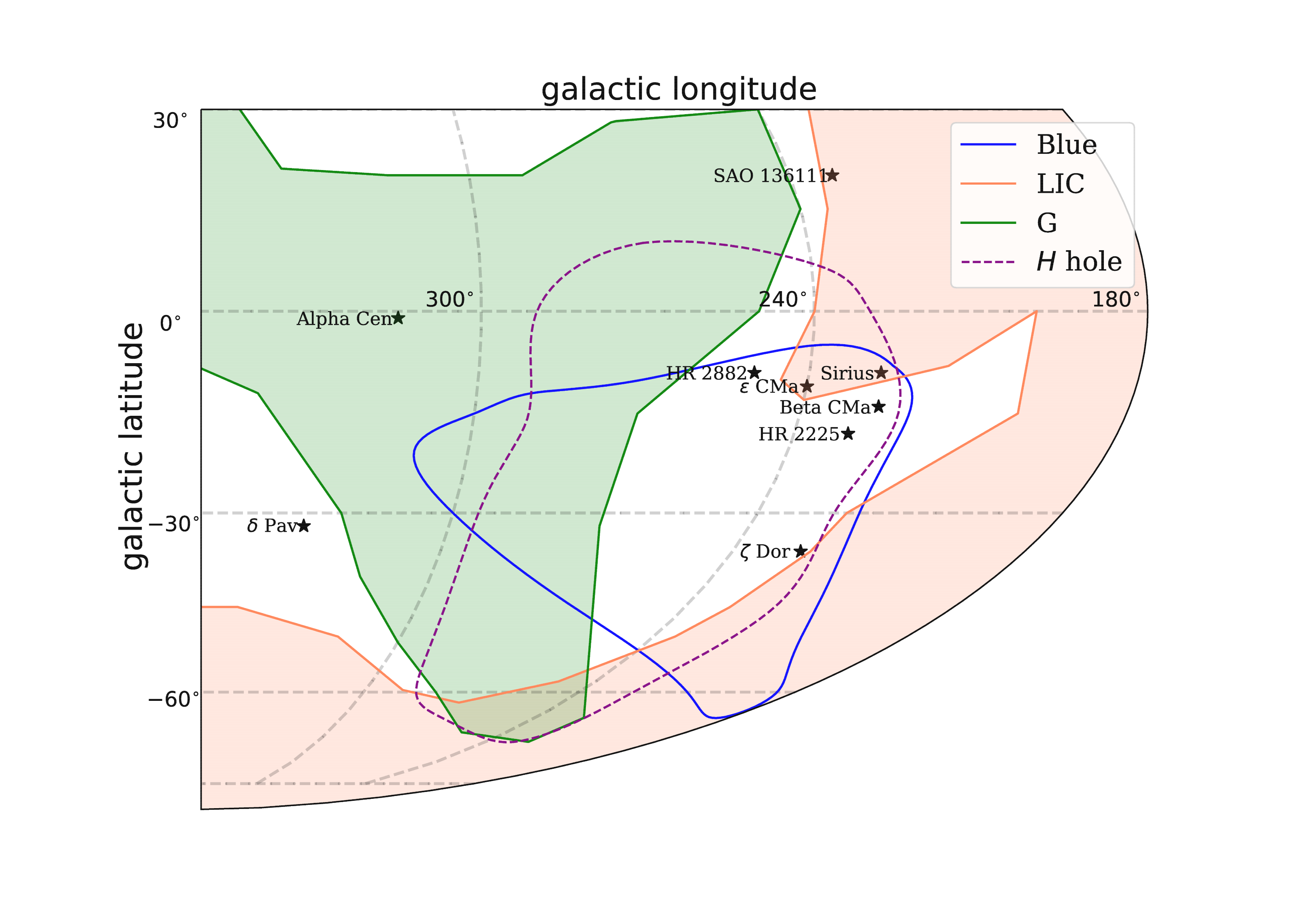}
\caption{Outer edges of the hydrogen hole (dashed line) and the Blue
  clouds (blue solid line) in Galactic
  coordinates. Also shown are the locations of the LIC (peach) and G (green)
  clouds. The plotted stars are those with individual clouds observed 
  in their lines of sight and (in many cases) knowledge of whether the
  solar hydrogen wall has been detected or not detected.
\label{fig:lism3}}
\end{figure}

In their study of the $\epsilon$~CMa line of sight with the Ech B and
G160M gratings on {\em HST}/GHRS, \cite{Gry1995} found absorption at the 
predicted radial velocities of the LIC and Blue clouds (their
components 1 and 2). The low value for the hydrogen column density
for component 1, log $N$(H~I)$\approx 17.34$, 
implies that the line of sight passes through the edge of the LIC
(see Figure~\ref{fig:lism3}), and the gas temperature $T=7450$~K is
similar to that found by \cite{Redfield2008} for the LIC. 
For the Blue cloud, the
inferred neutral hydrogen column density is also very small, 
log $N$(H~I)=16.88, and the
cloud gas temperature is low, $T=3600\pm 1500$~K, similar to that
found by \cite{Redfield2008}. The electron density in the Blue cloud,
$n_e=0.46^{+0.40}_{-0.30}$~cm$^{-3}$, found by \cite{Gry1995} could be
much larger than what they found for the LIC,
$n_e=0.09^{+0.23}_{-0.07}$~cm$^{-3}$. 
The high electron density of the Blue cloud, {\bf if real}, could
be explained either by a higher total density or less shielding from the EUV
radiation of $\epsilon$~CMa compared to the LIC or both effects. 
In addition to interstellar absorption in low excitation lines,
\cite{Gry1995} also found absorption features in the C~IV lines at the predicted
radial velocity of the LIC cloud. They could not rule out the
possibility that the C~IV absorption is stellar, perhaps from the
star's wind, rather than interstellar. Further study is needed to
determine whether the absorption indicates the presence of
highly ionized gas surrounding the cooler material in the LIC
and Blue clouds.

Interstellar absorption along the line of sight to $\beta$~CMa has
been studied by \cite{Dupin1998} using UV spectra from {\em HST}/GHRS 
and by \cite{Jenkins2000} using UV spectra from the {\em IMAPS} instrument. 
In their paper, \cite{Jenkins2000} identified a velocity component that
is at the
predicted radial velocity of the Blue cloud. There is also absorption
at the predicted radial velocity of the LIC, but \cite{Jenkins2000}
argued that this absorption is not likely from the LIC on the basis of the
unrealistically high ionization required to fit the observed absorption 
in many ions.   

\subsection{Stars near the edge of the hydrogen hole} 
Five stars are located near the edge of the hydrogen hole either just
inside or outside. Four show LIC absorption and one (Sirius) also shows
solar hydrogen wall absorption. At the outer edge of the hydrogen
hole, therefore, LIC gas is in contact with the outer heliosphere. 
For the lines of sight that also
show Blue cloud absorption, we place the Blue cloud just outside of the LIC. 

\subsection{Stars outside of but near the hydrogen hole} 
Outside of the hydrogen hole
  at Galactic longitudes greater than l=280$^{\circ}$, five of the six
  stars have solar hydrogen wall detections and all six have no detected
  absorption at radial velocities predicted by the LIC velocity
  vector. In these directions there must be neutral hydrogen gas flowing
  into the heliosphere to create the solar hydrogen wall, but this
  neutral gas has the velocity vector of the G or Aql cloud 
  rather than the LIC. 

\subsection{Comparison of the Blue cloud with the hydrogen hole}

Figure~\ref{fig:lism3} shows the outer contours of the hydrogen hole 
and the locations of $\epsilon$~CMa, $\beta$~CMa and some of the stars 
in Table~\ref{tab:outside} plotted in Galactic coordinates. 
Superimposed on the hydrogen hole is the boundary of the Blue cloud
based on Figure 3 in \cite{Redfield2008} and the stars listed in
Table~\ref{tab:outside}. The similar
morphologies of the hydrogen hole and the Blue cloud 
strongly imply their physical connection. 
The radial velocity of the Blue cloud in the direction of
$\epsilon$~CMa is 12.97~km~s$^{-1}$, which is blue shifted about 6
km~s$^{-1}$ relative to the predicted LIC velocity for this line of
sight. We interpret the Blue cloud as a
Str\"omgren shell that is driven towards the
heliosphere by excess pressure in the external H~II region. The
Blue cloud's lower temperature and higher gas density than
the LIC could result from compression of the Blue cloud gas leading to 
increased radiative cooling compared to the LIC.

We conclude from this analysis that in the hydrogen hole direction,
lines of sight from the Sun first pass through the Blue cloud
and then through ionized gas. Outside of the hydrogen hole 
at larger Galactic longitudes, the lines of sight from the Sun
pass through the G or other clouds rather than the LIC.
The stars lying at the edge of the hydrogen hole and
$\epsilon$~CMa have
lines of sight that first encounter a small column density of LIC gas.
The absence of LIC gas inside most of the hydrogen hole confirms that
the heliosphere lies at the edge of or just beyond the LIC.

%%%%%%%%%%%%%666666666666666666666666666666666666666666666666666666

\section{Is the LIC flow vector consistent with measurements of 
interstellar gas flowing through the heliosphere?}

\begin{figure}
\includegraphics[angle=0,width=7in]{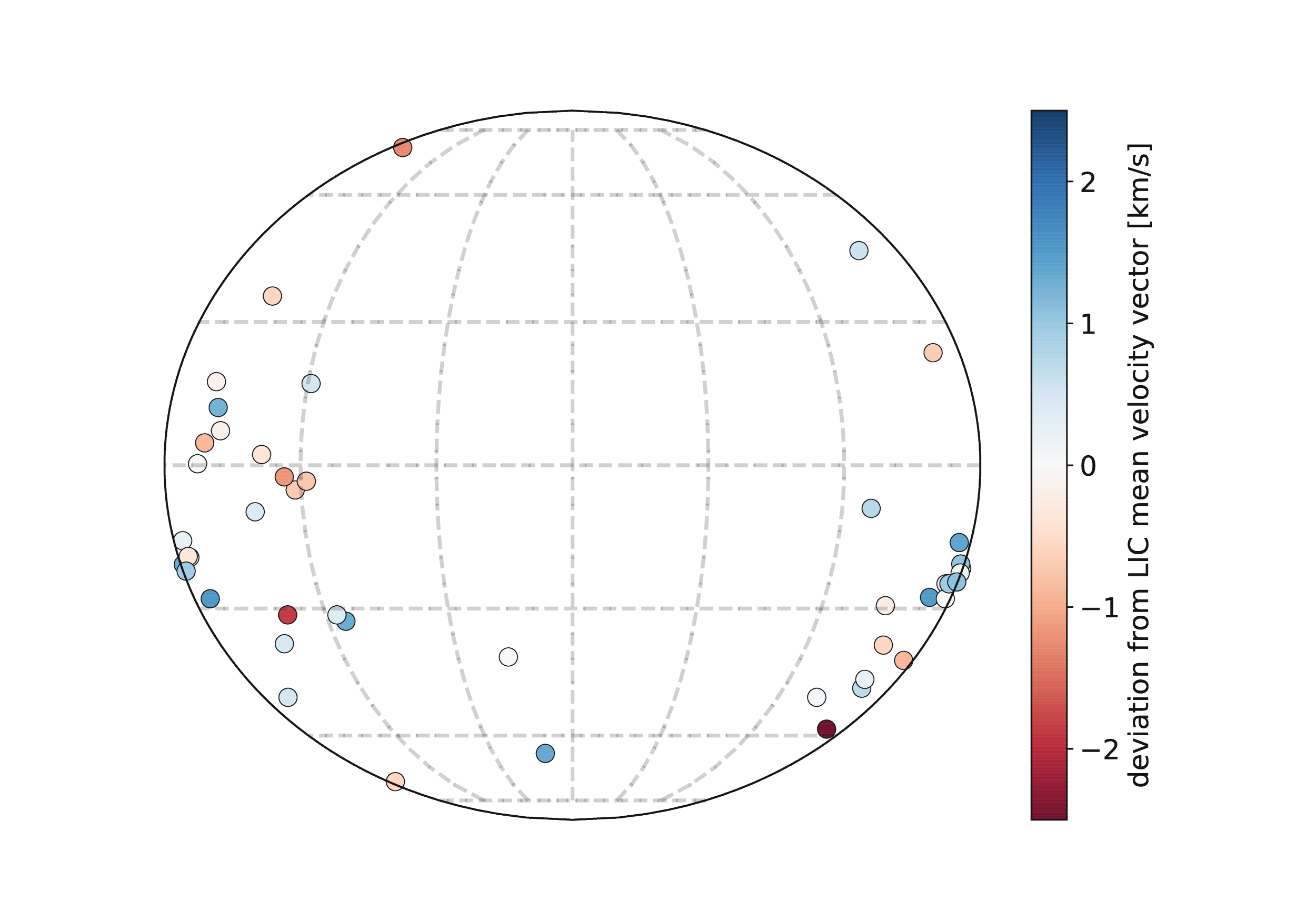}
\caption{Radial velocity deviations from the LIC mean velocity vector for LIC
  velocity components in the lines of sight to 45 stars. The data are
  plotted in Galactic Coordinates.}
\label{fig:lic2}
\end{figure}

In Section 1, we called attention to the flow vector for
neutral and singly ionized helium near the heliosphere 
inferred from observations with the {\em EUVE},
{\em IBEX}, {\em Ulysses}, and {\em STEREO} spacecraft. This vector,
which we call the ``inflow vector'', refers to gas just 
before entering the heliosphere where interactions with the Sun's 
gravity, radiation pressure, and solar wind particles
can alter the flow direction. The excellent agreement in speed, flow 
direction and temperature among these
different measurement techniques by different instruments provides 
the benchmark for the flow vector of
interstellar gas just outside of the heliosphere. The flow vector of
neutral hydrogen in the LIC cloud, which we call the ``LIC vector'', 
refers to gas located at one or a few parsecs from the
heliosphere where possible influences from the Sun and the solar wind
are negligible. Here we consider whether the inflow and LIC vectors
are in agreement, or whether they differ due
to some effect not yet taken into consideration. There are several
possibilities:
  
\begin{description}
\item[The inflow and LIC vectors agree within measurement errors]
The data in Table~\ref{tab:inflow} shows that the gas temperatures of
the two vectors are in
agreement, but the inflow speed of the LIC gas is 2--3~km~s$^{-1}$ too
slow, the ecliptic longitude of the LIC flow is about 3$^{\circ}$ too high,
and the ecliptic latitude about 2$^{\circ}$ too low. However, the difference 
in speed is only about $2.2\sigma$ and the difference in flow
directions only $1\sigma$. Given these small differences, one
could argue that the inflow and LIC flow vectors agree,
but new studies are needed to reduce measurement
errors and provide insight into the magnitude of potential systematic errors.

\item[The LIC flow may be inhomogeneous]
The difference in flow speeds of the two vectors, if real, could
test the kinematics of the LIC gas.
There is no physical reason why the low density gas in the
LIC should have homogeneous flow properties. In fact, the mean nonthermal
broadening of interstellar absorption lines in the LIC,
$\xi=1.62\pm 0.75$~km~s$^{-1}$ \citep{Redfield2008}, is nearly 
as large as the
difference in speed between the inflow and LIC vectors. Other nearby
interstellar clouds have values of $\xi$ as large as 3.6~km~s$^{-1}$,
about half of the cloud's sound speed. \cite{Gry2014} have proposed that the
multicloud scenario can be replaced by a single interstellar cloud with velocity
gradients, but \cite{Redfield2015} have argued that the data are
more accurately fit with multiple clouds each with its own velocity vector.

As a test for an inhomogeneous flow pattern in the LIC gas, we have selected
LIC velocity components that meet the following criteria: (a) the
uncertainty in the measured radial velocity is no more than three
times the precision of the instrumetal velocity scale 
(typically 1.5 km~s$^{-1}$), and (b) that the distance from
the Sun is no more than 4~pc. The latter criterion removes velocity
components with large uncertainties in $N$(H~I). Of the 62 LIC
velocity components, 45 meet both criteria. Figure~\ref{fig:lic2}
shows the deviations in the measured radial velocities from those
predicted by the LIC velocity vector.
The velocity deviations appear to be random, except for an excess of
blue points (positive radial velocities) near $l=180^{\circ}$ and 
$b=-25^{\circ}$. It is interesting that this location is close to the
tail of the LIC flow, $l=187.0^{\circ}\pm3.4^{\circ}$ and 
$b=-13.5^{\circ}\pm 3.3^{\circ}$ \citep{Redfield2008}. The reality of
this region of velocity deviation and a physical explanation for the 
similar direction as the tail of
the LIC flow requires additional data and investigation.  

\item[The LIC flow at it outer edge may differ from its mean value]
Outside of the hydrogen hole, a
thin region of LIC gas could be in contact with the ionized gas produced by 
radiation from $\epsilon$~CMa and Sirius. Contact with this ionized gas 
can alter the
flow of the adjacent LIC gas by a pressure difference
if the ionized gas has a higher or lower pressure than the LIC. 

\item[Could hydrogen and helium in the LIC have different flow vectors?]
Since the inflow vector is measured from neutral and ionized helium
and the LIC vector is measured from neutral hydrogen, we consider
whether hydrogen and helium could have different inflow vectors.
\cite{Lallement2005} compared the inflow direction of neutral helium
with that of neutral hydrogen observed from the glow of backscattered 
solar Lyman-$\alpha$ photons that were observed by the Solar Wind Anisotropies 
(SWAN) instrument on the {\em Solar and Heliospheric Observatory (SOHO)}.
The inflow directions of neutral helium and hydrogen differ by about
4$^{\circ}$ (see Table~\ref{tab:inflow}), which they explain as due to the 
inflowing hydrogen being a mixture
of pristine interstellar hydrogen and hydrogen atoms resulting from
charge exchange between solar wind electrons and interstellar protons. 
The properties of this hydrogen with mixed origins is very different
from the pristine interstellar helium observed by {\em IBEX, Ulysses,}
and {\em STEREO} and from the hydrogen observed at parsec distances by UV
spectroscopic measurements. Thus the question of whether there is a
difference between the flow
vectors of hydrogen and helium in the LIC prior to interactions in
the heliosphere remains open.

\item[Interstellar magnetic fields may be important]
Using {\em IBEX}
observations of the circular shaped ribbon of intense emission by
energetic neutral atoms, \cite{Zirnstein2016} derived the 
interstellar magnetic field strength of $2.93\pm 0.08$~$\mu$G outside
of the heliosphere. The magnetic pressure of this field is
comparable to the gas pressure in the heliopause leading to the
magnetic field draping and altering the shape of the heliopause region. Also,
the MHD calculations by \cite{Zank2013} show that for an interstellar 
magnetic field of this strength, the boundary between the outer 
heliopause and the
interstellar medium will be a bow wave rather than a shock wave.
At the center of the ribbon, the
direction of this magnetic field is 
$\lambda=227.28^{\circ}\pm 0.69^{\circ}$, $\beta=34.62^{\circ}\pm0.45^{\circ}$ in
ecliptic coordinates or $l=25.98^{\circ}\pm0.70^{\circ}$, $b=50.09^{\circ}\pm
0.57^{\circ}$ in Galactic coordinates. The 41 degree angle between the 
interstellar magnetic field and the inflow direction 
means that the magnetic field can change the apparent inflow direction
of ions relative to the undeflected neutrals.
The inflow direction of neutral hydrogen and helium atoms
is not influenced by the magnetic field unless these neutrals had
previously been ions before charge exchange. The consistent values 
of $\lambda$ and $\beta$ for He$^0$ measured by {\em IBEX} and {\em Ulysses} and
for He$^+$ measured by {\em STEREO} suggest that magnetic fields may
not play an important role in changing the inflow flow vector based on
ionized and neutral helium, but this is an open question.  

\end{description} 

We conclude that the inflow and LIC vectors are different and that this
difference indicates
that the heliosphere is now passing through a region different from the
main body of the LIC. The heliosphere may be inside the outer edge of
the LIC with properties modified by EUV radiation or contact with
ionized gas. Additional observations are needed to address this question.

%%%%%%%%%%%%%%%7777777777777777777777777777777777777777777777777

\section{Sightlines to Nearby Stars}

\begin{figure}
\includegraphics[angle=0,width=7.0in]{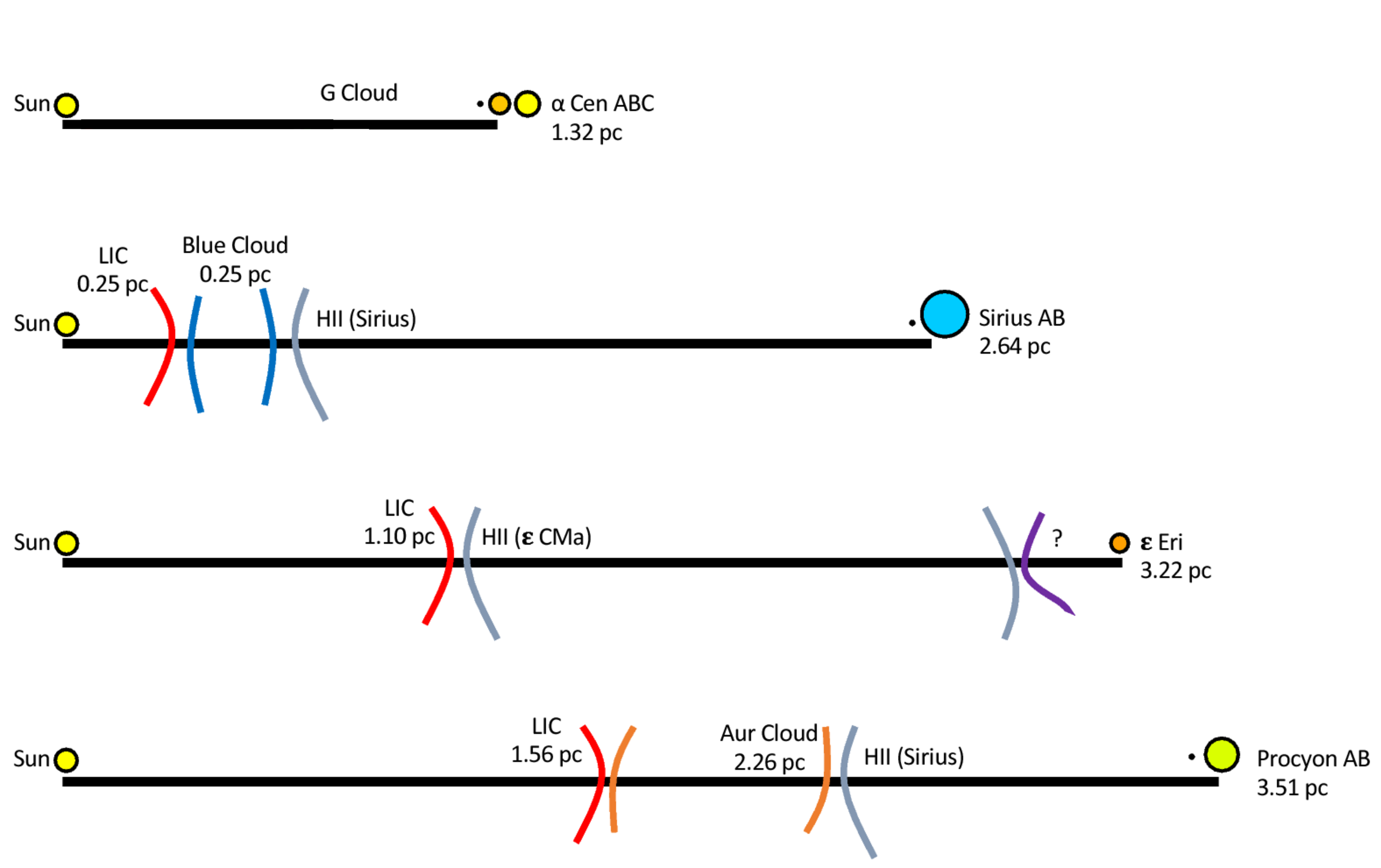}
\caption{Interstellar gas components along the lines of sight 
to four stars within 4 pc of the Sun. These lines of sight include 
contributions from the LIC, G, Blue, and Aur clouds and the H~II regions 
produced by EUV radiation from Sirius B and $\epsilon$~CMa.
\label{fig:InnerLISMfig}}
\end{figure}

\begin{figure}
\includegraphics[angle=0,width=7.0in]{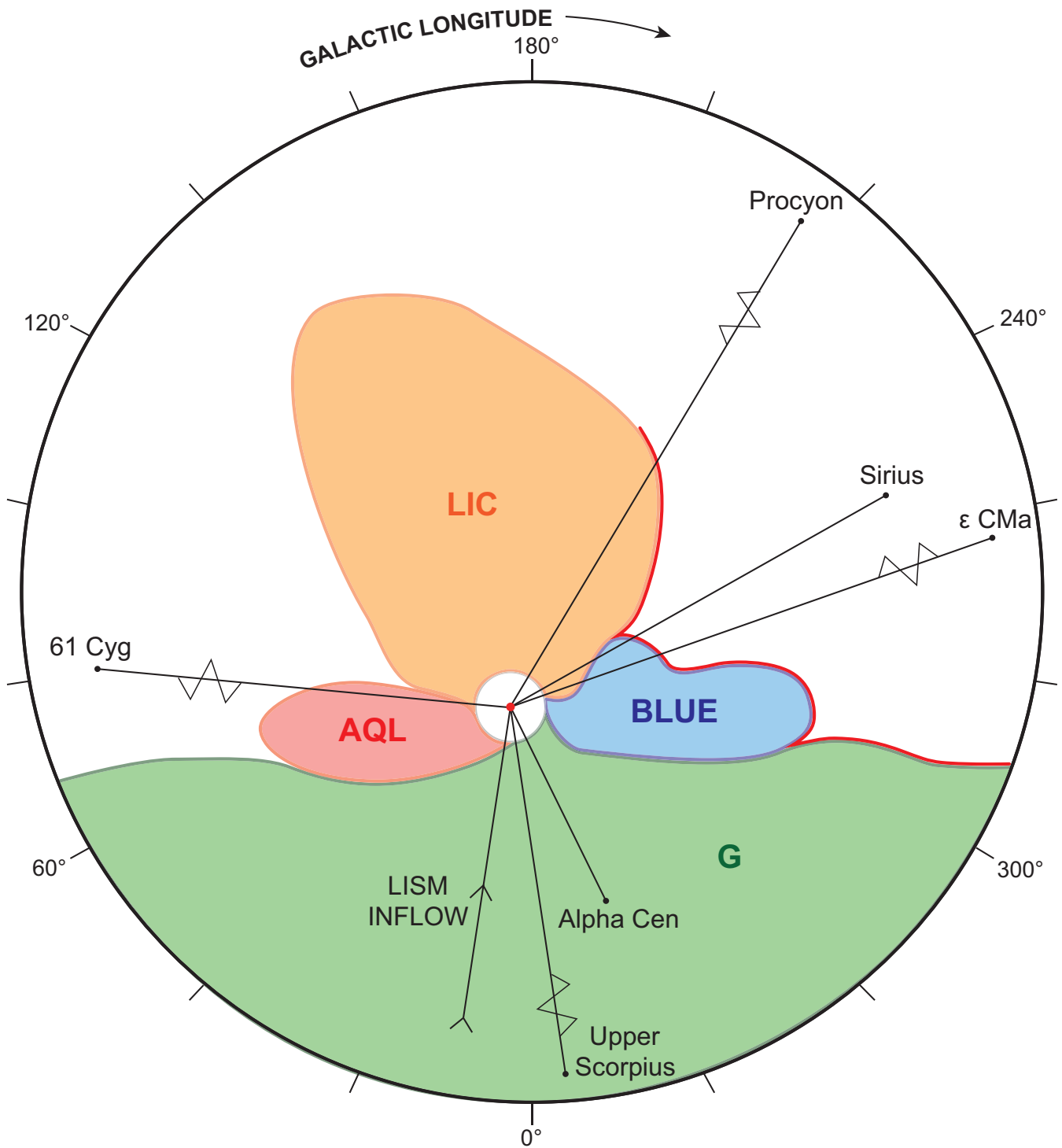}
\caption{The local ISM region within 3~pc of the Sun as viewed from the
  North Galactic pole showing the location of the four partially
  ionized clouds that are in contact with the outer heliosphere. Not
  shown are other clouds lying outside of the four clouds.
  Shown are the Sun (point), an
  exagerated representation of the heliopause (circle around the Sun)
  and the LIC, G, Aql, and Blue clouds. Lines of sight projected on to
  the Galactic equator are shown for 5 stars. Red shading shows the
  Str\"omgren shells produced by EUV radiation from $\epsilon$~CMa.
  Also shown are the
  direction of inflowing interstellar gas as seen from the Sun and the
  direction to the Upper Scorpius region of the Scorpius-Centauri
  Association where the most recent supernovae likely occured.  
\label{fig:interstellar_clouds_5}}
\end{figure}

We now describe the placement of warm interstellar clouds and H~II
intercloud gas along the lines of sight (LOS) to nearby stars. 
Figure~\ref{fig:InnerLISMfig} shows the gas components along the
sightlines to four nearby stars. The distance scales assume that 
$n$(H~I)=0.2~cm$^{-3}$ \citep{Slavin2008} for all of the 
partially ionized clouds. We
require that $\alpha$~Cen, $\epsilon$~Eri, and other stars with
detected astrospheric absorption are surrounded by
partially ionized hydrogen clouds. With these constraints, we propose 
that the sightlines to nearby stars have the following structures:

\begin{itemize}

\item {\bf The LOS to Sirius:} LIC and Blue are the two detected
partially ionized clouds in this LOS.  We propose that the 
LIC extends from the outer heliosphere to 0.25~pc in the direction of
Sirius, followed by the Blue cloud for a distance of 0.25~pc, and then 
H~II gas photoionized by Sirius~B
and $\epsilon$~CMa filling the remaining 2.14~pc to Sirius.

\item {\bf The LOS to $\alpha$~Cen:} Only the G cloud absorption is detected 
extending over a distance of 0.70~pc if we assume that
$n$(H~I) in the G cloud is same 
as in the LIC. The remaining 0.62~pc would filled with 
H~II gas from Sirius~B and $\epsilon$~CMa. 
Since $\alpha$~Cen has an astrosphere \citep{Wood2002}, it must by
surrounded by gas containing neutral hydrogen, which we assume is the
G cloud. However, heliospheric hydrogen wall absorption is detected
in the direction of $\alpha$~Cen \citep{Wood2005}, 
implying that the outer heliosphere must be in contact with neutral 
hydrogen gas in this direction. 
One explanation is to assume that a very
thin layer of the LIC provides the neutral
hydrogen needed to create the hydrogen wall in this direction but with
a hydrogen column density too small to be detected.
Another explanation is to extend the G
cloud all of the way to the outer heliosphere. This
can be accomplished by reducing the assumed $n$(H~I) in the G cloud to
0.11~cm$^{-3}$, half of the value of the LIC, but consistent with 
variability, the
precision of $n_e$ measurements in the LISM \citep{Redfield2008b}, and
with the measured $\log N$(H~I) = 17.6 to $\alpha$~Cen. 
Figure~\ref{fig:interstellar_clouds_5} shows this model. 

\item {\bf The LOS to $\epsilon$ Eri:} Spectra of $\epsilon$ Eri show 
absorption only at the LIC velocity, but the star has an astrosphere 
\citep{Wood2002}
and, therefore, must be located in a presently unknown cloud
containing neutral hydrogen. The LIC fills 
1.10~pc along this LOS leaving 
2.12~pc to be filled with ionized gas. Since Sirius is located 4.88~pc from 
$\epsilon$~Eri 
and 3.40~pc from the midpoint of the LOS from the Sun to $\epsilon$~Eri,
we do not expect that the H~II gas produced by Sirius~B can fill the 
missing 2.12~pc along the Sun-$\epsilon$~Eri LOS. Another hot white dwarf, 
40~Eri~B, is located
6.11~pc from $\epsilon$~Eri, but it is unlikely that its H~II region gas
comes close to the Sun-$\epsilon$~Eri LOS. Instead, $\epsilon$ CMa
is the likely source for this ionized gas. 
The gas-pressure of the
LIC, $(n_{HI}+n_e+n_p+n_{He})T=2710$~K~cm$^{-3}$, where 
$n_{HI}$=0.2 cm$^{-3}$, 
$n_e=n_p=0.07\pm 0.002$ cm$^{-3}$ \citep{Slavin2008}, 
$n_{He}/n_{H}=0.1$, and $T=7500$~K
\citep{Redfield2008}.
If there is gas pressure balance between the LIC and the H~II gas,
then the temperature of the ionized gas
along this LOS would be about 40,000~K.

\item {\bf The LOS to Procyon:} Absortion by the LIC and beyond it the
Aur cloud leave 1.25~pc of
path length to be filled with ionized gas. The morphology of the Aur
and other nearby clouds not in contact with the heliosphere will be
the subject of a forthcoming paper. Ionizing radiation from Sirius~B 
is the likely source of this ionized gas, since the separation of Sirius and 
Procyon is only 1.61~pc. This would be consistent with the gas 
temperature of about 7,500~K if there is gas-pressure balance between
the extended H~II region and the LIC.

\item {\bf The LOS to $\pi^1$~UMa, V368~Cep, MN~UMa, $\delta$~Dra, 47~Cas, and 
$\iota$~Cep:} Spectra of these stars centered near l=130$^{\circ}$, 
b=+30$^{\circ}$ with distances 14.4--35.4~pc all show interstellar
absorption only at the LIC velocity with no evidence for any other 
neutral gas in the LOS. Since the LIC lies in the immediate vicinity of the 
Sun, the remainder of these lines of sight must be ionized gas. In
addition to $\epsilon$~CMa,
GJ3753 (14.1~pc) and especially the hot white dwarf G191-B2B (59.9~pc) may be 
responsible for much of the ionizing radiation from this general direction. 

\item {\bf The LOS to 61~Vir, $\beta$~Com, $\tau$~Boo, and
$\chi$~Her:} Spectra of 
these stars, all located at high Galactic latitudes (b$>44^{\circ}$), 
show interstellar absorption only by the NGP cloud with 
no evidence for absorption by the LIC or any other cloud. Since the closest 
star, 61 Vir (d=8.53~pc), likely has a detected astrosphere \citep{Wood2005}, 
the NGP cloud must be located near to and in front of this star and perhaps 
the other stars. This leaves about 6.8~pc of the LOS to 
61 Vir and similar path lengths toward the other stars to be filled 
with ionized gas. The high Galactic latitude white dwarfs especially HZ~43 
but also GJ3753, GJ433.1, and UZ~Sex could provide part of this
ionizing radiation in addition to that provided by 
$\epsilon$~CMa and $\beta$~CMa.

\end{itemize}

Figure~\ref{fig:interstellar_clouds_5} is a schematic representation
of the four partially ionized clouds that are in contact with the
outer heliosphere as seen from the North Galactic pole. 
The figure shows the sight lines
to 5 stars projected on to a plane parallel to the Galactic equator, 
and the length of
each cloud located along each line of sight. Red shading indicates the 
Str\"omgren shells produced by EUV radiation from $\epsilon$~CMa. 
The direction of inflowing interstellar gas as seen from the Sun is
at $b\approx 20^{\circ}$ where the LIC, Aql, and G clouds may be in contact.

%%%%%%%%88888888888888888888888888888888888888888888888888888888888888

\section{\bf Could the LISM clouds in contact with the heliosphere be the
  source of $^{60}$Fe accretion?}

With a half-life of 2.6 Myr, the radioisotope $^{60}$Fe is 
produced during the very late evolution of massive stars and then ejected by
supernovae into the interstellar medium. The presence of this isotope in
deep ocean samples such as ferro-manganese crusts, nodules, and sediments
\citep{Knie1999,Wallner2016} and in the lunar regolith
\citep{Fimiani2016} indicates that supernovae have occured within the
last few Myr in the solar vicinity.
\cite{Wallner2016} found evidence for enhanced
$^{60}$Fe accretion during the time intervals 1.5--3.2 Myr and 6.5--8.7 Myr
ago but no evidence for $^{60}$Fe accretion above background
either more recently
or outside of these two time intervals. They concluded that the $^{60}$Fe
detected during these two time intervals was produced by one or more supernovae
occurring at these times.
The most likely location for these supernovae would be in the closest
association of massive stars, the Scorpius-Centaurus Association
at a distance of 100--150 pc. Within the Sco-Cen Association, 
the youngest
star forming region where the recent supernova likely occured is the
Upper Scorpius region centered at $l=352^{\circ}$ and
$b=-15^{\circ}$. \cite{Fry2015} concluded that the most likely 
explanation for the event timed at 2.2 My ago was the ejection
of $^{60}$Fe in the debris of an electron-capture supernova 
at a distance of about 100~pc
and the subsequent condensation of the $^{60}$Fe onto large 
($>0.2$~$\mu$m) interstellar grains. This could explain the 
amount of $^{60}$Fe that
arrived at Earth after traversing the interstellar medium and heliosphere.

Very recently, \cite{Koll2019} identified $^{60}$Fe in dust grains embedded
in Antarctic snow. After careful analysis, they concluded that the
$^{60}$Fe could not be explained by terrestrial nuclear explosions or by
cosmogenic sources, but instead must have a supernova origin. Unlike
the ocean core samples that were built up a long time ago,
the Antarctic snow sample is a very recent accumulation over the last 20 years.
The time scale for supernova debris to travel a distance of 100~pc
through the interstellar medium would be
about 200,000 yr \citep{Fry2015}, which is much less than the 
half-life of $^{60}$Fe or the time
of the most recent supernovae and thus cannot directly explain 
the very recent accretion of $^{60}$Fe.

Since iron is singly ionized in interstellar gas, iron ions flow
around rather than penetrating the heliosphere.
To reach the inner solar system, $^{60}$Fe must, therefore, be included
in interstellar grains. The large depletion of iron from the gas phase
of LISM clouds requires that most of the iron is resident in dust
grains that are likely olivene silicates \citep{Redfield2008,Frisch2011}.
{\em In situ} measurements of interstellar dust by
experiments on {\em Ulysses} and other spacecraft sample dust
grains with sizes larger than about 0.3 $\mu$m, because solar radiation
pressure and heliospheric magnetic fields filter out most of the
smaller grains \citep{Mann2010,Kruger2019}. 
Larger grains are expected to reach the inner
solar system without significant changes in direction or
speed. In their analysis of data from the {\em Ulysses} impact
ionization dust detectors, \cite{Strub2015} found that the
speed ($\approx 26$~km~s$^{-1}$) of large dust grains (sizes greater than 0.5
$\mu$m) and their flow
toward ecliptic longitude $\lambda=75^{\circ}\pm
30^{\circ}$ and latitude $\beta=-13^{\circ}\pm 4^{\circ}$
are similar to neutral helium gas in the LIC and other nearby clouds. 
The large uncertainty in $\lambda$ precludes identification of the dust
flow with the helium gas flow of a specific cloud, but the the data are
consistent with the dust flowing with the gas in the LIC, G, or other
nearby clouds. Interstellar grains with sizes less than 1$\mu$m
should be well coupled to the gas in warm clouds as the Larmor radius
for electrically changed grains is $<1$~pc for an interstellar magnetic
field of 5 $\mu$G \citep{Grun2000}.

We propose two possible explanations for the recent arrival of
interstellar dust grains containing $^{60}$Fe in
Antarctic snow. One is that the dust grains containing $^{60}$Fe are
resident in the warm LISM clouds and 
enter the heliosphere from one or all of the four clouds that are
in contact with the outer heliosphere. The density of dust grains in
the ionized intercloud medium should be much less than in the warm
clouds because strong UV radiation and shocks can destroy
grains and the low gas density means slower grain formation.
Since the Sun is moving
at a speed of 26.3~pc per million years through the cluster of local 
clouds, it entered the local cluster
about 200,000 years ago and will leave in about the same
time assuming that the warm clouds extend about 5~pc in all directions. This
is a rough estimate, but it gives a time scale for the input of dust 
grains containing $^{60}$Fe from a recent supernova. 
This scenario predicts continuous low level accretion of $^{60}$Fe
containing grains from warm clouds only 
when they are in contact with
the heliosphere. This scenario can be tested by searching for $^{60}$Fe
deeper in snow and ice fields going back to more than 200,000 years. 

An alternative explanation is that a large number of 
dust grains entered the heliosphere during the year 2005 when the
measured flux of dust increased a factor of 3 and the
inflow abruptly changed direction by $50^{\circ}\pm7^{\circ}$ \citep{Strub2015}.
This event might have been able to increase the terrestrial accretion 
rate to a level just
above background and thus appear as a one-time occurance. A possible
scenario would be a change in which cloud 
is feeding dust grains into the heliosphere. The absence of detected
$^{60}$Fe dust grains above background deeper in snow and ice fields would be
consistent with the recent detection being due to a one-time
event. 

New measurements are clearly needed to test between these two scenarios.
As suggested by \cite{Koll2019} and by \cite{Fry2015}, 
future measurements
of the $^{60}$Fe dust particles located deeper in Antarctic snow/ice fields
could provide a historical record of the Sun's motion through the LIC and other
clouds in the LISM.

\section{Conclusions and Further Work}

Observations and analysis of 62 sightlines with interstellar velocity
components consistent with the Local Interstellar Cloud  
vector permitted us to compute a three-dimensional model of the
LIC. This model extends from the heliosphere about 2~pc toward 
Galactic longitude $l=135^{\circ}$ and
latitude $b=+20^{\circ}$, but about 0~pc in the opposite direction 
($l=315^{\circ}$, $b=-20^{\circ}$)). This peculiar shape, which has
been identified
in previous studies, highlights the question of whether
the heliosphere is located inside or outside of the LIC. To better
understand this question, we analyzed spectroscopic data and obtained
the following results:

(1) As seen from the geometric center of the LIC, the distance to its edge
is less than 0.5~pc within a wide solid angle defined by 
$225^{\circ}\leq l \leq 290^{\circ}$ and  
$-60^{\circ}\leq b \leq+10^{\circ}$. We call this region of minimal neutral
hydrogen column density the ``hydrogen hole''. Inside of the hydrogen
hole are sight lines to the strongest source of EUV radiation
($\epsilon$~CMa), the second strongest source ($\beta$~CMa), and the
nearby hot white dwarf Sirius~B. Photoionization of neutral hydrogen
by the strong EUV radiation from these stars is
the most likely cause of the hydrogen hole.

(2) Inside of the hydrogen hole, sight lines to eight stars
show interstellar absorption by gas at the Blue cloud's radial 
velocity but not at the predicted LIC velocity. The outline of the Blue cloud 
overlaps that of the
hydrogen hole, indicating that the Blue cloud and the hydrogen hole are
probably physically associated. We propose that the outer 
edge of the Blue cloud is a Str\"omgren
shell being pushed against the outer heliosphere by higher ionized gas
behind it. The outer layers of other clouds facing
$\epsilon$~CMa are also Str\"omgren shells if they are not shielded by
other clouds. The presence of Si~III and likely C~IV absorption in
the sight lines to $\epsilon$~CMa and $\beta$~CMa support the argument
that these sight lines pass through Str\"omgren shells.
The radial velocity and higher gas pressure of the
Blue cloud are consistent with compression. Since the flow vector of
interstellar gas inside of the hydrogen hole differs substantally
from that of main body of the LIC, the heliosphere lies outside of the
LIC in this direction.

(3) The vector of interstellar gas flowing into the heliosphere as
measured by the {\em IBEX, Ulysses}, and {\em Stereo} spacecraft
differs from that inferred from interstellar absorption lines
representing the flow of LIC gas far from the heliosphere 
at a distance of roughly 1~pc. 
This difference of 2--3 km~s$^{-1}$ in speed and slightly different direction,
should be tested by precise new measurements. 
We conclude that the inflow and LIC vectors are different and 
propose that this difference indicates
that the heliosphere is now passing through a region different from the
main body of the LIC. The heliosphere could be inside the outer edge of
the LIC where the flow is modified by EUV radiation. 
Additional observations are needed to address this question.

(4) We propose a model for LISM immediately outside of the heliosphere.
The presence of heliospheric hydrogen wall absorption
in all directions requires that the outer heliosphere be in contact
with and be surrounded by interstellar gas containing a significant
amount of neutral hydrogen. In the hydrogen hole region, the Blue
cloud is in direct contact with the outer heliosphere. 
At the edge of the hydrogen hole, the
LIC is in contact with the outer heliosphere with the Blue cloud
lying immediately outside of the LIC.
Away from the hydrogen hole toward higher Galactic longitudes, the 
Aql cloud is in direct contact with the outer
heliosphere. In the direction of $\alpha$~Cen, there must be
partially ionized gas in contact with both the heliosphere and the
astrosphere. We adopt a model in which the G cloud fills
this entire line of sight to the star with neutral hydrogen density
$n$(H~I)$\approx 0.11$~cm$^{-3}$, although a very thin layer of LIC
gas may be in contact with the heliosphere in this direction. 
For $l=90^{\circ} - 235^{\circ}$,
the LIC is in contact with the outer heliosphere.
Our model with the heliosphere in direct
contact with four interstellar clouds 
may result from the directionality of the EUV radiation from $\epsilon$~CMa.
The different kinematics of the partially ionized interstellar
gas clouds may result from whether a cloud receives direct ionizing 
radiation or is
shielded by other clouds now and in the recent past. The complex
magnetic field surrounding the heliosphere may also play a role in
determining the shape and properties of these clouds.

(5) We describe the lines of sight to nearby stars in terms
of several partially ionized clouds and H~II gas which has been
photoionized by the EUV radiation from $\epsilon$~CMa and other stars.
The modest degree of ionization in the nearby intercloud gas and the strong EUV
radiation field suggest that the intercloud gas is
an irregularly shaped Str\"omgren sphere rather than a recombining
plasma following a supernova shock. 
 
(6) Finally, we note that the heliosphere is leaving a region of
space where the LIC, G, Aql, and Blue clouds are located. We propose
that the very recent measurement in Antarctic snow of enhanced
$^{60}$Fe from the debris of a supernova could be explained by 
the inflow of interstellar grains 
containing $^{60}$Fe from the warm clouds in contact with the heliosphere 
either continuously at a low level or during an unusual event.

Our models for the LIC morphology and the very local ISM are updates 
of our previous studies \citep{Redfield2000,Redfield2008}. 
In a subsequent paper,
we will present the results of a new three-dimensional LISM model
including data studied by \cite{Malamut2014} and more recently
observed sightlines. When
this is available, we will reexamine the extent to which stellar EUV
radiation may explain the properties of the intercloud gas between the
LISM.

\acknowledgements

We acknowledge support through the NASA HST Grant GO-11568 from the
Space Telescope Science Institute, which is operated by the
Association of Universities for Research in Astronomy, Inc. for NASA,
under contract NAS 5-26555.
Support for {\it HST} observing programs \#11568 was provided by NASA
through a grant from the Space Telescope Science Institute.  We thank
John Vallerga for a very thoughtful referee report, 
Martin Barstow for computing the ionizing flux from Sirius~B, and
Steven Burrows for his graphics. 
JLL thanks the Erwin Schr\"odinger International Institute for
Mathematics and Physics at the University of Vienna 
for their hospitality and opportunity to learn
about nucleosynthetic isotope anomalies. Our
research has made use of NASA's Astrophysics Data System Bibliographic
Services and the SIMBAD database, operated at CDS, Strasbourg, France.

{\it Facilities:} {HST (GHRS, STIS), FUSE, EUVE, CHIPS, ULYSSES}

\clearpage
%Figure 1

\clearpage
%Figure 2

\clearpage
%Figure 3

\clearpage
%Figure 4

\clearpage
%Figure 5

\clearpage
%Figure 6

\clearpage

\clearpage
%Figure 8

\clearpage
%Figure 9

\clearpage
%Figure 10

\clearpage
%Figure 11
\clearpage

%Table inflow (1)
\begin{deluxetable}{lllllll}
\tablewidth{0pt}
\tabletypesize{\small}
\tablecaption{Heliospheric and astronomical measurements of
  inflowing LISM gas\label{tab:inflow}}
\tablehead{Source & Method & $v_{\rm LISM}$ (km~s$^{-1}$) & $T$ (K) & 
$\lambda$ (ecliptic) & $\beta$ (ecliptic) & Ref.\tablenotemark{a}\\}
\startdata
{\it EUVE} & He$^0$ & $24.5\pm 2$ & $6500\pm 2000$ & $74.7\pm0.5$ &
$-5.7\pm0.5$ & 1\\
{\it IBEX} & He$^0$ & 26 & $8710^{+440}_{-680}$ & $75.75\pm0.92$ & 
$-5.29\pm0.05$ & 2\\
{\it IBEX} &  He$^0$ & $27.0\pm1.3$ & & $74.5\pm1.7$ & $-5.2\pm0.3$ & 
3\\
{\it IBEX} &  He$^0$ & $25.76\pm0.4$ & $7440\pm260$ & $75.75\pm0.5$ & 
$-5.16\pm0.10$ & 4\\
{\it IBEX} &  He$^0$ & $26.21\pm0.37$ & $7691\pm230$ & $75.41\pm0.40$
& $-5.03 \pm0.07$ & 5\\ 
{\it Ulysses} & He$^0$ & $26.08\pm 0.21$ & $7260\pm270$ & $75.54\pm0.19$ &
$-5.44\pm0.24$ & 6\\
{\it STEREO} & He$^+$ PUI & & & $75.21\pm0.04$ & & 7\\
{\it STEREO} & He$^+$ PUI & & & $75.41\pm0.34$ & & 8\\ \hline
LIC cloud & H$^0$ & $23.84\pm0.90$ & $7500\pm1300$ & $78.53\pm3.4$ &
$-7.20\pm3.3$ & 9\\
LIC cloud & H$^0$ & 23.9 & & 78.6 & --7.0 & 10\\
G cloud & H$^0$ & $29.6\pm1.1$ & $5500\pm400$ & $71.11\pm1.9$ & $-8.52\pm3.6$
& 9\\
%Aur cloud & H$^0$ & $25.22\pm0.81$ & & 87.46 & --30.02 & 9\\
Blue cloud & H$^0$ & $13.89\pm0.89$ & $3900\pm2300$ & 79.09 & --26.60 & 9\\ 
\hline
{\it SWAN} & H$^0$ & $22\pm 1$ & $11500\pm 1000$ & 72.5 & --8.8 & 11\\
{\it Ulysses} & dust & 26 & & $75\pm 30$ & $-13\pm 4$ & 12\\
\hline
\enddata
\tablenotetext{a}{(1) \cite{Vallerga2004};
(2) \cite{Mobius2015a}; (3) \cite{Leonard2015};
(4) \cite{Bzowski2015}; (5) \cite{Swaczyna2018};
(6) \cite{Wood2015}; (7) \cite{Mobius2015b};
(8) \cite{Taut2018}; (9) \cite{Redfield2008}; (10) This paper; 
(11) \cite{Lallement2005}; (12) \cite{Strub2015}.} 
\end{deluxetable}
\clearpage

%From Seth
%Table licmem (2)
\startlongtable
\begin{deluxetable}{lcccccccccccc}
\tablewidth{0pt}
\tabletypesize{\tiny}
\tablecaption{LIC Cloud Sight Line Properties \label{tab:licmem}}
\tablehead{Star & HD & $l$ & $b$ & $d$ & $v$ & $\log N({\rm Ion})$ & Ion & 
Reference & $\log N({\rm H~I})$ & $d^{\rm edge}_{\rm obs}$ & 
$d^{\rm edge}_{\rm model}$ & $\sigma$\tablenotemark{a} \\
Name & \# & (deg) & (deg) & (pc) & (\kms) & (cm$^{-2}$) & & & (cm$^{-2}$) & 
(pc) & (pc) & \\}
 \startdata
$\alpha$ CMa A & 48915  & 227.2 & --8.9 &  2.64 & $ 18.55$ & $12.43 \pm 0.17 $ 
& DI   & 1   & $17.24  \pm 0.17 $ & $0.28  \pm 0.11 $ & 0.24   & 0.4\\
$\alpha$ CMa B & 48915B & 227.2 & --8.9 &  2.64 & $ 17.60$ & $12.81 \pm 0.10 $ 
& DI   & 2   & $17.62  \pm 0.10 $ & $0.67  \pm 0.16 $ & 0.24   & 2.8\\
$\epsilon$ Eri & 22049  & 195.8 & --48.1 &  3.22 & $ 18.73$ & $13.03\pm 0.06 $ 
& DI   & 3   & $17.837 \pm 0.061$ & $1.11  \pm 0.16 $ & 0.91   & 1.3\\
 $\alpha$ CMi  & 61421  & 213.7 & 13.0 &  3.51 & $ 19.76$ & $13.08 \pm 0.04 $ 
& DI   & 4   & $17.887 \pm 0.042$ & $1.25  \pm 0.12 $ & 0.38   & 7.3\\
$\epsilon$ Ind & 209100 & 336.2 & --48.0 &  3.62 & $ -9.20$ & $13.20\pm 0.10 $ 
& DI   & 5   & $18.01  \pm 0.05 $ & $1.65  \pm 0.38 $ & 0.13   & 4.0\\
   $\tau$ Cet  & 10700  & 173.1 & --73.4 &  3.65 & $ 12.34$ & $13.182\pm 0.002$
& DI   & 6   & $17.989 \pm 0.011$ & $1.58  \pm 0.32 $ & 1.18   & 1.2\\
     40 Eri A  & 26965  & 200.8 & --38.1 &  4.98 & $ 21.73$ & $13.03 \pm 0.1 $ 
& DI   & 7   & $17.84  \pm 0.10 $ & $1.11  \pm 0.26 $ & 0.79  & 1.3\\
 $\eta$ Cas A  & 4614   & 122.6 & --5.1 &  5.95 & $ 11.18$ & $12.43 \pm 0.09 $ 
& FeII & 8   & $18.10  \pm 0.14 $ & $2.04  \pm 0.67 $ & 2.63   & --0.9\\
  $\alpha$ Lyr & 172167 & 67.5 & 19.2 &  7.68 & $-12.90$ & $13.03 \pm 0.1  $ 
& FeII & 9  & $18.70  \pm 0.15 $ & $8.1   \pm 2.8$ & 0.63 & 2.7 \\
  $\alpha$ PsA & 216956 & 20.5 & --64.9 &  7.70 & $ -5.87$ & $12.58 \pm 0.1  $ 
& FeII & 10   & $18.25  \pm 0.15 $ & $2.9   \pm 1.0  $ & 0.57   & 2.3\\
 $\chi^1$ Ori  & 39587  & 188.5 & --2.7 &  8.66 & $ 23.08$ & $12.98 \pm 0.01 $ 
& DI   & 8   & $17.787 \pm 0.015$ & $0.992 \pm 0.034$ & 1.35   & --10.5\\
 $\delta$ Eri  & 23249  & 198.1 & --46.0 &  9.04 & $ 19.60$ & $13.056\pm 0.007$
& DI   & 6   & $17.863 \pm 0.013$ & $1.18  \pm 0.24 $ & 0.86   & 1.3\\
       CF UMa  & 103095 & 168.5 & 73.8 &  9.09 & $  2.05$ & $12.58 \pm 0.14 $ 
& MgII & 8   & $18.02  \pm 0.28 $ & $1.7   \pm 1.1  $ & 0.32   & 1.3\\
$\kappa^1$ Cet & 20630  & 178.2 & --43.1 &  9.14 & $ 20.84$ & $12.68\pm 0.13 $ 
& DI   & 8   & $17.49  \pm 0.13 $ & $0.50  \pm 0.15 $ & 1.36   & --5.8\\
       EP Eri  & 17925  & 192.1 & --58.3 & 10.4 & $ 19.50$ & $13.13\pm 0.10 $ 
& DI   & 11  & $17.94  \pm 0.10 $ & $1.40  \pm 0.32 $ & 1.00   & 1.3 \\
  $\beta$ Gem  & 62509  & 192.2 & 23.4 & 10.4 & $ 19.65$ & $13.20\pm 0.10 $ 
& DI   & 3   & $18.01  \pm 0.10 $ & $1.65  \pm 0.38 $ & 1.08   & 1.5\\
       13 Per  & 16895  & 141.2 & --9.6 & 11.1 & $ 16.45$ & $12.61\pm 0.02 $ 
& FeII & 8   & $18.28  \pm 0.11 $ & $3.09  \pm 0.81 $ & 3.00   & 0.1\\
 $\gamma$ Ser  & 142860 & 27.7 & 45.7 & 11.3 & $-18.19$ & $12.33\pm 0.10 $ 
& FeII & 8   & $18.00  \pm 0.15 $ & $1.62  \pm 0.56 $ & 0.13   & 2.7\\
       HR 1925 & 37394  & 158.4 & 11.9 & 12.3 & $ 17.50$ & $13.43\pm 0.005$ 
& DI   & 6   & $18.239 \pm 0.012$ & $2.81  \pm 0.56 $ & 2.45   & 0.6\\
 $\alpha$ Aur  & 34029  & 162.6 & 4.6 & 13.1 & $ 21.48$ & $13.44\pm 0.02 $ 
& DI   & 4   & $18.247 \pm 0.023$ & $2.86  \pm 0.15 $ & 2.50   & 2.4\\
         HR 8  & 166    & 111.3 & --32.8 & 13.7 & $  6.50$ & $13.466\pm 0.006$ 
& DI   & 6   & $18.273 \pm 0.013$ & $3.04  \pm 0.61 $ & 2.51   & 0.9\\
       72 Her  & 157214 & 55.9 & 32.3 & 14.3 & $-15.51$ & $14.32\pm 0.44 $ 
& FeII & 8   & $19.99  \pm 0.45 $ & $160   \pm 170$\tablenotemark{b,c} & 0.36 
& \nodata \\
 $\pi^1$ UMa  & 72905 & 150.6 & 35.7 & 14.4 & $12.49$ & $13.33\pm 0.02 $ & DI 
& 12  & $18.137  \pm 0.023 $ & $2.22 \pm 0.12$ & 1.96 & 2.2 \\
    V1119 Tau & 35296 & 187.2 & --10.3 & 14.4 & $23.47$ & $12.80\pm 0.14 $ 
& DI & 13  & $17.61 \pm 0.14 $ & $0.66 \pm 0.21$ & 1.38 & --3.4 \\
       99 Her  & 165908 & 57.0 & 22.3 & 15.6 & $-17.43$ & $13.27\pm 0.14 $ 
& FeII & 8   & $18.94  \pm 0.18 $ & $14.1  \pm 5.8$\tablenotemark{b} & 0.43  
& \nodata \\
 $\sigma$ Boo  & 128167 & 45.6 & 67.2 & 15.8 & $ -2.58$ & $11.82\pm 0.14 $ 
& FeII & 14  & $17.49  \pm 0.18 $ & $0.50  \pm 0.21 $ & 0.18   & 1.5\\
 $\sigma$ Boo  & 128167 & 45.6 & 67.2 & 15.8 & $ -2.28$ & $11.23\pm 0.10 $ 
& MgII & 14  & $16.67  \pm 0.27 $ & $0.076  \pm 0.047 $\tablenotemark{d} 
& 0.18  & \nodata \\
  $\beta$ Cas  & 432    & 117.5 & --3.3 & 16.8 & $  9.15$ & $13.38\pm 0.08 $ 
& DI   & 3,15 & $18.187 \pm 0.081$ & $2.49  \pm 0.46 $ & 2.42   & 0.2\\
 $\tau^6$ Eri  & 23754  & 217.4 & --50.3 & 17.6 & $ 16.99$ & $12.71\pm 0.12 $ 
& FeII & 14  & $18.38  \pm 0.16 $ & $3.9   \pm 1.5  $ & 0.56   & 2.2\\
 $\tau^6$ Eri  & 23754  & 217.4 & --50.3 & 17.6 & $ 16.99$ & $13.01\pm 0.08 $ 
& MgII & 14  & $18.45  \pm 0.26 $ & $4.6   \pm 2.7  $\tablenotemark{d} & 0.56 
& \nodata\\
       DX Leo  & 82443  & 201.2 & 46.1 & 17.8 & $ 11.00$ & $12.88\pm 0.15 $ 
& DI   & 11  & $17.69  \pm 0.15 $ & $0.79  \pm 0.28 $ & 0.44   & 1.3\\
     V368 Cep  & 220140 & 118.5 & 16.9 & 19.2 & $  6.00$ & $13.13\pm 0.10 $ 
& DI   & 11  & $17.94  \pm 0.05 $ & $1.40  \pm 0.32 $ & 1.81   & --1.3\\
  $\alpha$ Tri & 11443  & 138.6 & --31.4 & 19.4 & $ 17.89$ & $13.30\pm 0.10 $ 
& DI   & 3   & $18.11  \pm 0.10 $ & $2.07  \pm 0.48 $ & 3.12   & --2.2\\
      HR 4345  & 97334  & 184.3 & 67.3 & 21.9 & $  4.30$ & $13.00\pm 0.01 $ 
& DI   & 6   & $17.807 \pm 0.015$ & $1.04  \pm 0.21 $ & 0.35   & 3.3\\
   SAO 136111  & 73350  & 232.1 & 20.0 & 24.0 & $ 12.00$ & $13.339\pm 0.009$ 
& DI   & 6   & $18.146 \pm 0.014$ & $2.27  \pm 0.45 $ & 0.15   & 4.7\\
 $\lambda$ And & 222107 & 109.9 & --14.5 & 26.4 & $  6.50$ & $13.68\pm 0.15$ 
& DI   & 5   & $18.49  \pm 0.15 $ & $4.97  \pm 1.72 $ & 2.07   & 1.7\\
 $\lambda$ And & 222107 & 109.9 & --14.5 & 26.4 & $  4.52$ & $12.92\pm 0.26 $ 
& FeII   & 16   & $18.59  \pm 0.28 $ & $6.3  \pm 4.1 $ & 2.32   & 1.0 \\
 $\lambda$ And & 222107 & 109.9 & --14.5 & 26.4 & $  4.97$ & $13.17\pm 0.22 $ 
& MgII   & 16   & $18.61  \pm 0.33 $ & $6.6  \pm 5.0 $\tablenotemark{d} 
& 2.32   & \nodata \\
 $\sigma$ Cet  & 15798  & 191.1 & --63.8 & 26.7 & $ 15.99$ & $13.19\pm 0.10 $ 
& FeII & 8   & $18.86  \pm 0.15 $ & $11.7  \pm 4.1$\tablenotemark{b} &  
& \nodata \\
       HR 860  & 17948  & 137.2 & 2.2 & 26.7 & $ 15.10$ & $12.60\pm 0.03 $ 
& FeII & 8   & $18.27  \pm 0.12 $ & $3.02  \pm 0.80 $ & 2.76   & 0.3\\
    SAO 32862  & 198084 & 94.4 & 9.1 & 27.3 & $ -2.60$ & $12.80\pm 0.02 $ 
& FeII & 8   & $18.47  \pm 0.11 $ & $4.8  \pm 1.3 $ & 1.28   & 2.8\\
   $\eta$ Ari  & 13555  & 147.1 & --37.8 & 28.9 & $ 16.99$ & $12.38\pm 0.05 $ 
& FeII & 8   & $18.05  \pm 0.12 $ & $1.82  \pm 0.51 $ & 3.03   & --2.4\\
       PW And  & 1405   & 114.6 & --31.4 & 29.0\tablenotemark{e} & $  8.50$ 
& $13.23  \pm 0.10 $ & DI   & 11  & $18.04  \pm 0.10 $ & $1.76  \pm 0.41 $ 
& 2.64   & --2.1\\
    SAO 85045  & 157466 & 47.5 & 29.8 & 29.3 & $-19.02$ & $13.70\pm 0.36 $ 
& FeII & 8   & $19.37  \pm 0.38 $ & $38    \pm 33$\tablenotemark{b} & 0.27 
& \nodata \\
    SAO 85045  & 157466 & 47.5 & 29.8 & 29.3 & $-19.02$ & $14.15\pm 0.59 $ 
& MgII & 8   & $19.59  \pm 0.64 $ & $63    \pm 93$\tablenotemark{b,c,d} 
&  0.27 & \nodata\\
 $\delta$ Cas  & 8538   & 127.2 & --2.4 & 30.5 & $ 13.05$ & $12.795\pm 0.1  $ 
& MgII &10,17& $18.24  \pm 0.27 $ & $2.8   \pm 1.7  $ & 2.69   & 0.1\\
      HR 1099  & 22468  & 184.9 & --41.6 & 30.7 & $ 21.90$ & $13.06\pm 0.03 $ 
& DI   & 15  & $17.867 \pm 0.032$ & $1.193 \pm 0.088$ & 1.16   & 0.4\\
  $\alpha$ Gru & 209952 & 350.0 & --52.5 & 31.0 & $-10.93$ & $12.98\pm 0.11 $ 
& FeII & 14  & $18.65  \pm 0.16 $ & $7.2   \pm 2.6$ & 0.18 & 2.7 \\
  $\alpha$ Gru & 209952 & 350.0 & --52.5 & 31.0 & $-10.93$ & $13.79\pm 0.22 $ 
& MgII & 14  & $19.23  \pm 0.33 $ & $28   \pm 21$\tablenotemark{b,d} & 0.18 
& \nodata \\
       DK UMa  & 82210  & 142.6 & 38.9 & 31.9 & $  9.41$ & $13.14\pm 0.01 $ 
& DI   & 8   & $17.947 \pm 0.015$ & $1.434 \pm 0.049$ & 1.78   & --7.0\\
$\epsilon$ Gru & 215789 & 338.3 & --56.5 & 39.5 & $ -7.30$ & $12.86\pm 0.1  $ 
& FeII & 18  & $18.53  \pm 0.15 $ & $5.5   \pm 1.9  $ & 0.18   & 2.8\\
$\epsilon$ Gru & 215789 & 338.3 & --56.5 & 39.5 & $ -7.30$ & $13.95\pm 0.1  $ 
& MgII & 18  & $19.39  \pm 0.27 $ & $40   \pm 24  $\tablenotemark{b,d} & 0.18  
& \nodata \\
      101 Tau  & 31845  & 185.1 & --16.0 & 40.8 & $ 22.40$ & $12.49\pm 0.05 $ 
& MgII & 19  & $17.93  \pm 0.25 $ & $1.38  \pm 0.80 $ & 1.46   & --0.1\\
    SAO 93981  & 28568  & 180.5 & --21.4 & 41.5 & $ 23.90$ & $13.21\pm 0.02 $ 
& DI   & 6   & $18.017 \pm 0.023$ & $1.685 \pm 0.089$ & 1.72   & --0.4\\
   SAO 111879  & 28736  & 190.2 & --27.6 & 43.5 & $ 21.60$ & $12.65\pm 0.05 $ 
& MgII & 19  & $18.09  \pm 0.25 $ & $2.0   \pm 1.2  $ & 1.08   & 0.8\\
     V471 Tau  & \nodata& 172.5 & --27.9 & 44.1 & $ 20.90$ & $13.382\pm 0.007$ 
& DI   & 6   & $18.189 \pm 0.013$ & $2.50  \pm 0.50 $ & 2.20   & 0.6\\
    SAO 76593  & 27808  & 174.8 & --19.1 & 44.3 & $ 23.10$ & $12.86\pm 0.11 $ 
& MgII & 19  & $18.30  \pm 0.27 $ & $3.2   \pm 2.0  $ & 2.14   & 0.5\\
    SAO 76683  & 29419  & 176.0 & --15.6 & 44.5 & $ 23.20$ & $12.69\pm 0.08 $ 
& MgII & 19  & $18.13  \pm 0.26 $ & $2.2   \pm 1.3  $ & 2.09   & 0.1\\
    SAO 93982  & 28608  & 185.1 & --24.7 & 44.7 & $ 23.20$ & $12.42\pm 0.06 $ 
& MgII & 19  & $17.86  \pm 0.25 $ & $1.17  \pm 0.69 $ & 1.33   & --0.2\\
    SAO 93831  & 26784  & 182.4 & --27.9 & 45.0 & $ 23.00$ & $12.32\pm 0.13 $ 
& MgII & 19  & $17.76  \pm 0.28 $ & $0.93  \pm 0.60 $ & 1.45   & --0.9\\
    SAO 94033  & 29225  & 181.6 & --20.5 & 45.7 & $ 22.50$ & $12.68\pm 0.04 $ 
& MgII & 19  & $18.12  \pm 0.25 $ & $2.1   \pm 1.2  $ & 1.65   & 0.4\\
    SAO 93963  & 28406  & 178.8 & --20.6 & 46.4 & $ 22.10$ & $12.67\pm 0.07 $ 
& MgII & 19  & $18.11  \pm 0.26 $ & $2.1   \pm 1.2  $ & 1.86   & 0.2\\
    SAO 93945  & 28237  & 183.7 & --24.7 & 46.4 & $ 22.40$ & $12.42\pm 0.12 $ 
& MgII & 19  & $17.86  \pm 0.27 $ & $1.17  \pm 0.74 $ & 1.42   & --0.3\\
     V993 Tau  & 28205  & 180.4 & --22.4 & 47.0 & $ 23.30$ & $13.18\pm 0.01 $ 
& DI   & 6   & $17.987 \pm 0.015$ & $1.572 \pm 0.054$ & 1.72   & --2.7\\
    SAO 76609  & 28033  & 175.4 & --18.9 & 49.4 & $ 23.60$ & $13.33\pm 0.02 $ 
& DI   & 6   & $18.137 \pm 0.023$ & $2.22  \pm 0.12 $ & 2.11  & 0.9\\
    SAO 56530  & 21847  & 156.2 & --16.6 & 49.5 & $ 21.10$ & $13.14\pm 0.37 $ 
& MgII & 19  & $18.58  \pm 0.44 $ & $6.2   \pm 6.3$\tablenotemark{c} & 2.90 
& \nodata \\
    SAO 93885  & 27561  & 180.4 & --24.3 & 52.4 & $ 22.20$ & $12.50\pm 0.06 $ 
& MgII & 19  & $17.94  \pm 0.25 $ & $1.41  \pm 0.83 $ & 1.68   & --0.3\\
    SAO 93913  & 27848  & 178.6 & --22.0 & 53.5 & $ 22.40$ & $12.56\pm 0.24 $ 
& MgII & 19  & $18.00  \pm 0.34 $ & $1.6   \pm 1.3  $ & 1.86   & --0.2\\
      HR 1608  & 32008  & 209.6 & --29.4 & 54.0 & $ 21.60$ & $13.02\pm 0.05 $ 
& DI   & 6   & $17.827 \pm 0.051$ & $1.09  \pm 0.22 $ & 0.59   & 2.3\\
       45 Aur  & 43905  & 161.1 & 17.3 & 59.2 & $ 18.47$ & $12.47\pm 0.03 $ 
& FeII & 8   & $18.14  \pm 0.12 $ & $2.24  \pm 0.60 $ & 2.29   & --0.1\\
     G191-B2B  & \nodata& 156.0 & 7.1 & 59.9 & $ 19.19$ & $13.36\pm 0.03 $ 
& DI   & 8   & $18.167 \pm 0.032$ & $2.38  \pm 0.18 $ & 2.59   & --1.2\\
  $\iota$ Cap  & 203387 & 33.6 & --40.8 & 60.3 & $-12.06$ & $13.11\pm 0.33 $ 
& DI   & 8   & $17.92  \pm 0.33 $ & $1.3   \pm 1.0  $ & 0.50   & 0.8\\
    $\eta$ Aur & 32630  & 165.4 & 0.3 & 74.6 & $21.5$ & $12.27\pm 0.08 $ 
& FeII   & 20  & $17.94  \pm 0.14 $ & $1.41  \pm 0.45 $ &  2.48  & --2.4\\
     Feige 24  & \nodata& 166.0 & --50.3 & 91.7 & $ 17.60$ & $13.19\pm 0.14 $ 
& DI   & 21  & $18.00  \pm 0.14 $ & $1.61  \pm 0.52 $ & 1.67   & --0.1\\
\enddata
\tablenotetext{a}{$(d^{\rm edge}_{\rm obs} - 
d^{\rm edge}_{\rm model})/\sigma(d^{\rm edge}_{\rm obs})$}
\tablenotetext{b}{Sight line ignored because observed value is $>$5$\sigma$ 
from median value.}
\tablenotetext{c}{Sight line ignored because relative error 
($\sigma(d^{\rm edge}_{\rm obs})/d^{\rm edge}_{\rm obs}$) is of order unity 
(i.e., $>$0.9).}
\tablenotetext{d}{This measurement is not used in the subsequent analysis.  
Preference is given instead to the complementary Fe\,II measurement.}
\tablenotetext{e}{\citet{metchev09}}
\tablerefs{(1) \citet{bertin95}; (2) \citet{hebrard99}; (3) \citet{dring97}; 
(4) \citet{linsky95}; (5) \citet{wood96}; (6) \citet{Wood2005}; 
(7) \citet{wood98}; (8) \citet{Redfield2004a}; (9) \citet{lallement95}; 
(10) \citet{ferlet95}; (11) \citet{wood00}; (12) \citet{woodpi1uma14}; 
(13) \citet{wood14}; (14) \citet{Redfield2002}; (15) \citet{piskunov97}; 
(16) \citet{Malamut2014}; (17) \citet{lallement97}; 
(18) \citet{lecavelierdesetangs97}; (19) \citet{Redfield2001}; 
(20) \citet{welsh10pasp}; (21) \citet{vennes00}.}
\end{deluxetable}  
\clearpage

%Table licfit (3)
\begin{deluxetable}{lc}
\tabletypesize{\normalsize}
\tablecaption{LIC Model Parameters \label{tab:licfit}}
\tablewidth{0pt}
\tablehead{
Property & Parameter}
\startdata
Sight lines used  & 62 \\
$x_{\rm center}$ (pc) & --0.8 pc \\
$y_{\rm center}$ (pc) & +0.7 pc  \\
$z_{\rm center}$ (pc) & --0.4 pc \\
reduced $\chi^2$ & 8.3 \\
Median difference (pc) & 0.40 \\
\hline 
\multicolumn{2}{c}{Spherical Harmonic coefficients} \\ 
\hline
$a_{\rm 0,\phantom{-}0}$ &  +4.708 \\
$a_{\rm 1,\phantom{-}0}$ & --0.519 \\
$a_{\rm 1,-1}$           &  +0.421 \\
$a_{\rm 1,+1}$ &  --0.524 \\
$a_{\rm 2,\phantom{-}0}$ &  +0.197 \\
$a_{\rm 2,-1}$           & --0.360 \\
$a_{\rm 2,+1}$ & +0.172 \\
$a_{\rm 2,-2}$           & --0.193 \\
$a_{\rm 2,+2}$ &  +0.022 \\
\enddata
\end{deluxetable}
\clearpage

%Table composition (4)
\begin{table}
\caption{Composition of Gas Along Selected Lines of Sight 
\label{tab:composition}}
\begin{center}
\begin{tabular}{llllllll}
\hline\hline
Star & $l$ & $b$ & $d$(pc) & Cloud$^{c}$ & log[$N$(H~I)]$^{c}$ & 
$\Delta d$(neutral)$^{d}$ & $\Delta d$(ionized)$^{d}$\\
\hline
$\alpha$~Cen$^{a}$ & 315.7 & --0.9 & 1.32 & G & 17.6 & 0.70 & 0.62\\

Sirius A & 227.2 & --8.9 & 2.64 & LIC & 17.2 & 0.25 & \\
         &       &       &      & Blue & 17.2 & 0.25 & \\
         &       &       &      & Sum &       & 0.50 & 2.14\\

$\epsilon$ Eri$^{a}$ & 5.8 & --48.1 & 3.22 & LIC & 17.8 & 1.10 & 
2.12\\

61 Cyg$^{a}$ & 82.3 & --5.8 & 3.49 & Aql & 17.8 & 1.10 & \\
       &      &       &      & Eri & 17.8 & 1.10 & \\
       &      &       &      & Sum &      & 2.20 & 1.29\\

Procyon & 213.7 & +13.0 & 3.51 & LIC & 17.9 & 1.56 & \\
        &       &       &      & Aur & 17.6 & 0.70 & \\
        &       &       &      & Sum &      & 2.26 & 1.25\\

$\epsilon$ Ind$^{a}$ & 336.2 & --48.0 & 3.63 & LIC & 16.6 & 0.14 & 
3.49\\

61 Vir$^{a}$ & 311.9 & +44.1 & 8.53 & NGP & (18.0) & (1.7) & (6.8)\\

$\beta$ Com & 43.5 & +85.4 & 9.15 & NGP & (18.0) & (1.7) & (7.4) \\

$\pi^1$ UMa & 150.6 & +35.7 & 14.6 & LIC & 17.85 & 2.30 & 12.3 \\

$\tau$ Boo & 358.9 & +73.9 & 15.6 & NGP & (18.0) & (1.7) & 13.9\\

51 Peg & 90.1 & --34.7& 15.6 & Eri & (17.9) & (1.6) & \\
       &      &       &      & Hya & (17.4) & (0.4) & \\
       &      &       &      & Sum &        & (2.0) & (13.6)\\

$\chi$ Her$^{b}$ & 67.7 & +50.3 & 15.9 & NGP & (18.0) & (1.7) & 
14.2\\
\hline
\end{tabular}
\end{center}
$^a$ Astrosphere detected by \cite{Wood2005} indicates that a 
cloud with neutral hydrogen must surround the star.

$^b$No astrosphere detected by \cite{Wood2005}.

$^{c}$ Cloud identification and hydrogen column densities from 
\cite{Redfield2008}.

$^{d}$ Distance (pc) to the star through neutral and ionized gas.
\end{table}
\clearpage

%Table shells (5)
\begin{table}
\caption{Are clouds in and near the hydrogen hole Str\"omgren shells? 
\label{shells}}
\begin{center}
\begin{tabular}{lllll}
\hline\hline
Line of sight & log $N$(H~I) & cloud & thickness (pc) & property\\ 
through  & & & & \\
\hline
Str\"omgren shell& 17.10 & & 0.2 & for $n$(H~I)=0.2 cm$^{-3}$\\
$\epsilon$~CMa & 16.76 & Blue & 0.12 & shell (1/e EUV photons absorbed)\\
$\epsilon$~CMa & 17.34 & LIC & 0.35 & most EUV photons absorbed\\
Sirius         & 17.2  & Blue & 0.25 & shell (1/e photons absorbed)\\
Sirius         & 17.2  & LIC & 0.25 & shell (1/e photons absorbed)\\
$\zeta$~Dor    & 17.8  & Blue & 1.0 & all EUV photons absorbed\\
HR 2225        & 17.9  & Blue & 1.3 & all EUV photons absorbed\\
\hline\hline
\end{tabular}
\end{center}
\end{table}
\clearpage

%Table outside (6)
\begin{table}
\caption{Stars located Inside and Near the Hydrogen Hole \label{tab:outside}}
\begin{center}
\begin{tabular}{llccccccc}
\hline\hline
Star & HD & d & l & b & Solar & LIC & Other & Hydrogen\\
     & & (pc) & & & H wall & detection & clouds & hole\\ 
\hline

$\zeta$ Dor & 33262 & 11.7 & 226 & --36 & N & N & Blue, Dor & inside\\
HR~2225 & 43162 & 16.7 & 230 & -18 & N & N & Blue & inside\\
$\beta$~Pic & 39060 & 19.3 & 258 & --31 & -- & N & Blue & inside\\
HR~2882 & 59967 & 21.8 & 250 & --9 & N & N & Blue & inside\\ 
$\beta$~Car & 80007 & 34.1 & 286 & --14 & -- & N & Blue, G & inside\\
HR~2265 & 43940 & 61.9 & 245 & --22 & -- & N & Blue, Dor & inside\\
WD 0621-376 & -- & 78 & 245 & --21 & -- & N & Blue, Dor & inside\\
$\alpha$~Car & 45348 & 96 & 261 & --25 & -- & N & Blue & inside\\
$\epsilon$~CMa & 52089 & 124 & 239.8 & --11.4 & -- & Y & Blue, other & inside\\ 
\hline
Sirius & 48915 & 2.6 & 227 & --9 & Y & Y & Blue & edge\\ 
EP~Eri & 17925 & 10.4 & 192 & --58 & -- & Y & Blue & edge\\
SAO~136111 & 73350 & 23.6 & 232 & +20 & N & Y & --- & edge\\ 
$\sigma$~Cet & 15798 & 25.8 & 191 & --64 & -- & Y & Blue, G & edge\\
$\beta$~CMa & 44743 & 151 & 226.1 & --14.3 & -- & N & Blue, other & edge\\
\hline
$\alpha$~Cen & 128620 & 1.3 & 315 & --1 & Y & N & G & outside\\
70~Oph~A & 165341 & 5.1 & 30 & +11 & Y & N & G, Aql, Oph & outside\\
36 Oph & 155886 & 6.0 & 358 & +7 & Y & N & G & outside\\
$\delta$~Pav & 190248 & 6.1 & 329 & --32 & Y & N & Vel, Dor & outside\\
HR~6748 & 165185 & 17.4 & 356 & --7 & Y & N & Aql & outside\\ 
$\rho$~Oct & 137333 & 66 & 307 & --23 & -- & N &  Blue, G & outside\\
\hline\hline
\end{tabular}
\end{center}
\end{table}

\end{document}